\documentclass[aps,prl,twocolumn,superscriptaddress,longbibliography]{revtex4-2}
\usepackage{amsmath,amssymb}
\usepackage{graphicx,color}
\usepackage{times}
\usepackage[pdftex,colorlinks,breaklinks]{hyperref}
\hypersetup{linkcolor=blue,citecolor=blue,filecolor=dullmagenta,urlcolor=blue, 
 pdftitle={Spectral and eigenvector statistics for complex symmetric random matrices},
 pdfauthor={Akemann, Fyodorov, Savin}, pdflang={English} }

\newcommand{\aver}[1]{\left\langle #1 \right\rangle}
\newcommand{\daver}[1]{\left\langle\!\left\langle #1\right\rangle\!\right\rangle}
\renewcommand{\Im}{\textrm{Im}}
\renewcommand{\Re}{\textrm{Re}}
\renewcommand{\vec}[1]{\boldsymbol{#1}}
\newcommand{\vv}{|\vec{v}^T\vec{v}|^2}
\newcommand{\Tr}{\mathrm{Tr}}
\newcommand{\id}{\mathbb{I}}
\begin{document}
\title{Spectral Density and Eigenvector Nonorthogonality in Complex Symmetric Random Matrices}
\author{Gernot Akemann}
 \affiliation{Faculty of Physics, Bielefeld University, P.O. Box 100131, 33501, Bielefeld, Germany}
\author{Yan V. Fyodorov}
\affiliation{Faculty of Physics, Bielefeld University, P.O. Box 100131, 33501, Bielefeld, Germany}
 \affiliation{Department of Mathematics, King’s College London, London, WC2R 2LS, United Kingdom}
\author{Dmitry V. Savin}
 \affiliation{Department of Mathematics, Brunel University of London, Uxbridge, UB8 3PH, United Kingdom}
\date{\today}
\begin{abstract}
Non-Hermitian random matrices with statistical spectral characteristics beyond the standard Ginibre ensembles have recently emerged in the description of dissipative quantum many-body systems as well as in non-ergodic wave transport in complex media. We investigate the class AI$^\dag$ of complex symmetric random matrices, for which available analytic results remain scarce. Using a recently proposed framework by one of the authors, we analyze this class for Gaussian entries and derive an explicit, closed-form expression for the joint distribution of a complex eigenvalue and its right eigenvector for arbitrary matrix size $N\ge 2$ in the entire complex plane. From this, we obtain the distribution of the eigenvector non-orthogonality overlap  and the mean eigenvalue density, both for finite $N$ and in the large-$N$ limit. Notably, at the spectral edge both the eigenvalue density and eigenvector statistics exhibits a limiting behavior that differs from the Ginibre universtality class. This behavior is expected to be universal, as further supported by numerical evidence for Bernoulli random matrices.
\end{abstract}
\makeatletter\let\MyTitle\@title\makeatother 
\maketitle

Non-Hermitian effects arising in quantum many-body or optical wave systems due to inherent mechanisms of dissipation, loss or gain have attracted vigorous research interest in the last decade, giving rise to an entirely new field, 
see Ref.~\cite{ashida2020non} for a review. In this context it was conjectured long ago  \cite{grobe1988quantum} that the theory of non-Hermitian random matrices provides a natural reference framework for identifying, characterizing, and quantifying universal phenomena brought about by non-Hermiticity to quantum systems dominated by effects of chaos and disorder. In recent years the utility of such an approach has been further supported by extensive studies of systems displaying many-body quantum chaos and eventually many-body localization phenomena \cite{akemann2019universal,denisov2019universal,jaiswal2019universality,can2019random,hamazaki2019non,sa2020spectrala,sa2020spectralb,sa2020complex,wang2020hierarchy,li2021spectral,tarnowski2021random,lange2021random,garcia2022symmetry,kulkarni2022lindbladian,ghosh2022spectral,suthar2022non,cipolloni2023entanglement,sa2023symmetry,de2022non,de2023non,ghosh2023eigenvector,orgad2024dynamical,jisha2024universality,richter2025integrability,almeida2025universality,rufo2025quantum,villasenor2025correspondence,wold2025experimental,chirame2025open}. In particular, numerical evidences combined with heuristic arguments \cite{hamazaki2020universality} suggest that, among the 38 possible non-Hermitian random matrix symmetry classes \cite{Kawabata2019,Bernard2002,Magnea2008}, there are only three types of local spectral correlations in the spectral bulk. This conjecture was recently extended to include the spectral edge behavior \cite{akemann2025complex}. The simplest representatives for these three  universality classes are complex Ginibre (class A), complex symmetric (class AI$^\dag$), and complex self-dual (class AII$^{\dag}$) Gaussian random matrices. While Ginibre matrices have been thoroughly investigated analytically for decades \cite{byun2025progress}, the analysis of the two non-standard classes remained predominantly confined to numerical simulations. In this way it has been observed, e.g., that their eigenvalue spacing distributions in the bulk are well approximated by a two-dimensional Coulomb gas at noninteger inverse temperatures \cite{akemann2022spacing,akemann2025two}.

This Letter aims to advance our understanding of the symmetry class AI$^\dag$ by providing exact analytical results for both spectral and eigenvector statistics of complex symmetric Gaussian random matrices.  Matrices of this type appear in diverse physical contexts, including $SU(2)$ gauge theory with imaginary chemical potential and fermions in pseudo-real representation \cite{kanazawa2021new}, XXZ spin chains with imaginary disorder \cite{akemann2025two}, variants of non-Hermitian Sachdev-Ye-Kitaev model \cite{garcia2022symmetry}, Liouville operators of dissipative quantum spin chains and Lindbladian superoperators \cite{hamazaki2020universality,sa2023symmetry},
non-Hermitian Anderson localization \cite{huang2020spectral}, scattering matrix pole statistics in quantum chaotic systems \cite{sommers1999s}, and non-ergodic wave transport in complex media \cite{prado2025nonergodic}.  Despite its relevance, this class has resisted standard random-matrix techniques, and only a handful of analytical results exists \cite{akemann2025complex,forrester2025dualities,kulkarni25}. In particular, even such fundamental spectral characteristics as the 
eigenvalue density remains unknown for general $N\times N$ complex symmetric matrices beyond the special limiting case of complex symmetric matrices remaining parametrically close to their real symmetric counterparts as $N\to \infty$ \cite{sommers1999s}, and the spacing distribution for $N=2$ \cite{hamazaki2020universality,jaiswal2019universality}.

To analyze these systems, it is essential to recall a key algebraic property: any generic non-Hermitian random matrix $J$ is non-normal, satisfying $J^{\dagger}J\ne JJ^{\dagger}$, yet remains diagonalizable. This gives rise to two distinct eigenproblems:
\begin{equation}\label{eig_prob}
  J\vec{v}_i = z_i\vec{v}_i\quad\mbox{and}\quad
  \vec{u}_i^{\dagger} J = z_i\vec{u}_i^{\dagger}\,,
\end{equation}
where $\vec{v}_i$ and $\vec{u}_i$ denote the right and left eigenvectors, respectively. Unlike in the Hermitian case, these eigenvectors are not orthogonal but instead form a biorthogonal system: $\vec{u}^{\dag}_k\vec{v}_l=0$ for $k\ne l$. The degree of non-orthogonality plays a central role in many applications and is most conveniently quantified by the so-called overlap matrix $\mathcal{O}$, with entries $\mathcal{O}_{kl}:=(\vec{u}^{\dag}_k\vec{u}_l)(\vec{v}^{\dag}_l\vec{v}_k)$, under the normalization $\vec{u}^{\dag}_k\vec{v}_k=1$.
In such a context the deviation $\mathcal{O}-\id$ from the identity serves as a natural indicator of non-normality of the matrix $J$.  Moreover, the diagonal entries $\mathcal{O}_{kk}\ge 1$ can be shown to control the spectral sensitivity of $J$ to small additive perturbations, with $\kappa=\sqrt{\mathcal{O}_{kk}}$ known in numerical analysis as the eigenvalue condition number \cite{trefethen2020spectra}. The same entries play an important role in quantum wave scattering, where they are referred to as Petermann factors \cite{schomerus2000quantum}. The matrix ${\cal O}$ features in quantum decay laws \cite{savi97} and influences transient dynamics also in classical systems \cite{ChalkerMehlig1998}, including their relaxation, entropy production and stability \cite{ChalkerMehlig2000,burd14,erdos2018power,Tarnowski_2020,Gudowska_neuro,Fyodorov_entropyproduction_2025}. It manifests itself in scattering, transport and other properties of single-particle as well as many-body quantum systems \cite{fyodorovSavin2012,schomerus2000quantum,savi97,gros14,davy2019probing,fyodorov2022eigenfunction,cipolloni2023non,cipolloni2023entan,ghosh2023eigenvector,chirame2025open,bao2025initial}. This broad relevance has motivated a long-standing interest in the statistics of the entries $\mathcal{O}_{kl}$ for random linear operators and matrices. This line of research was initiated by the seminal works of Chalker and Mehlig \cite{ChalkerMehlig1998,ChalkerMehlig2000}, who studied the mean nonorthogonality factors in the limit of large $N\gg 1$, a topic that has remained highly active ever since \cite{janik1999correlations,fyodorovmehlig2002,poli09b,burda2014dysonian,walters2015note,belinschi2017squared,bourgadedoubach2020,fyodorov2018CMP,Akemann2020quatern,dubach2021a,WFC1,akemann2020determinantal,akemann2020appa,akemann2020universal,dubach2021b,fyodorov2021condition,osman2024universality,cipolloni2024optimala,cipolloni2025optimal,tarnowski2024condition,zhang2024mean,wurfel2024mean,crumpton2025mean,akemann2025pfaffian,noda2025determinantal}.


\textit{Main results}.---In this Letter we derive exact results for the joint probability density function (JPDF) $\mathcal{P}(z,\vec{v})$ of a complex eigenvalue $z$ and the associated right eigenvector $\vec{v}$ of a random complex symmetric matrix $J=X+iY$, where $X$ and $Y$ are two independent copies of the Gaussian orthogonal ensemble (GOE) of size $N$. The latter is characterized by real Gaussian-distributed entries with mean zero and the covariances $\aver{X_{nm}X_{kl}}=N^{-1}(\delta_{nk}\delta_{ml}+\delta_{nl}\delta_{mk})$, and similarly for $Y$, where $\aver{\cdots}$ denotes the ensemble average.

Working with complex symmetric matrices we found it more natural to normalize the left and right eigenvectors as $\vec{v}^\dagger\vec{v}=\vec{u}^\dagger\vec{u}=1$ rather than $\vec{u}^\dagger\vec{v}=1$. It is important to note that due to the symmetry $J^T=J$ the corresponding left eigenvectors are related to the right ones as $\vec{u}^*=\vec{v}$, where the star denotes complex conjugation. This enables us to express the associated diagonal entry of the overlap matrix solely in terms of $\vec{v}$, which in our normalization becomes
\begin{equation}\label{nonorth}
\mathcal{O}_{\mathrm{diag}} = \frac{(\vec{v}^\dagger\vec{v})(\vec{u}^\dagger\vec{u})}{|\vec{u}^\dagger\vec{v}|^2}
  = \frac{1}{|\vec{v}^T\vec{v}|^2} \equiv 1+t,
\end{equation}
where the parameter $t\geq0$ quantifies the degree of eigenvector nonorthogonality. In particular, this implies that for complex symmetric matrices the knowledge of $\mathcal{P}(z,\vec{v})$ provides the full description of both the spectral density and the nonorthogonality statistics in the entire complex plane.

To evaluate such an object we follow the general framework for non-Hermitian random matrices recently developed by one of us \cite{fyodorov2025kac}, which extends the Kac-Rice approach to complex eigenvalue problems. We begin with the following representation \cite{sommers1994eigenvector,fyodorov2025kac}:
\begin{equation}\label{jpdf-def}
  \mathcal{P}(z,\vec{v}) = \frac{\delta(\vec{v}^\dagger\vec{v}-1)}{\pi N}
  \aver{\delta^{(2)}[(J-z)\vec{v}]\left|\frac{d}{dz}\det(J-z)\right|^2}\,.
\end{equation}
 Here $\delta^{(2)}(\vec{v})=\prod_{i=1}^{N}\delta(\mathrm{Re\,}v_i)\delta(\mathrm{Im\,}v_i)$ is the delta-function of a complex-valued vector argument, $\vec{v}=(v_1,\ldots,v_N)^T$. The statistical averaging over the ensemble can be performed exactly using a variant of the supersymmetry technique tailored to this problem, as detailed in the Supplemental Material (SM) \cite{SM}. In this way one arrives at an exact integral representation for the JPDF valid for all $N\geq2$,
\begin{align}\label{jpdf-int}
  \mathcal{P}(z,\vec{v}) &= \frac{\delta(\vec{v}^\dagger\vec{v}-1)N^{2N+1}}{8\pi}
  \int_0^{\infty}\!dq\,q^3e^{-\frac{N}{2}q^2}
  \int\!\frac{d^2\vec{k}}{(2\pi)^{2N}}  \nonumber\\
  &\times e^{-\frac{N}{2}\mathcal{L}_{\vec{k}}(z,\vec{v})}
  \frac{\partial^2}{\partial{z}\,\partial{z}^*} (q^2+|z|^2)^{N-4} D_q(z,\vec{v},\vec{k})\,,
\end{align}
with $\mathcal{L}_{\vec{k}}(z,\vec{v}) = \vec{k}^\dagger\vec{k}+|\vec{k}^\dagger\vec{v}|^2 + z(\vec{k}^T\vec{v})-z^*(\vec{k}^T\vec{v})^*$ acting as an effective Lagrangian governing the integration over the complex $N$-dimensional vector field $\vec{k}$ and $D_q(z,\vec{v},\vec{k})$ being
\begin{equation}\label{Det}
  D_q(z,\vec{v},\vec{k}) =
  \det\left[(q^2+|z|^2)\id_4
  - \begin{pmatrix}zA & qB\\ qB^* & -z^*A^*\end{pmatrix}\right]\,.
\end{equation}
Here, the $2\times2$ matrices $A$ and $B$ are given by
\begin{equation}\label{Det_AB}\nonumber
  A = \begin{pmatrix} \vec{k}^T\vec{v} & \vec{v}^T\vec{v} \\
        \vec{k}^T\vec{k} & \vec{v}^T\vec{k} \end{pmatrix}\,,\quad
  B = \begin{pmatrix} \vec{k}^\dagger\vec{v} & 1 \\
        \vec{k}^\dagger\vec{k} & \vec{v}^\dagger\vec{k} \end{pmatrix}\,.
\end{equation}

The advantage of the obtained representation lies in the quadratic dependence of the Lagrangian $\mathcal{L}_{\vec{k}}(z,\vec{v})$ on the complex vector $\vec{k}$. This structure enables an exact evaluation of the $\vec{k}$-integration in Eq.~(\ref{jpdf-int}) via Wick's theorem, as the integrand is a polynomial in $\vec{k}$ in the pre-exponential factor. While algebraically intensive, the computation can be efficiently handled using the method developed in the SM (see Sec.~\ref{SM2} of \cite{SM} for details). The final result takes the following form:
\begin{equation}\label{jpdf-exact}
  \mathcal{P}(z,\vec{v}) =
  \frac{\delta(\vec{v}^\dagger\vec{v}-1)}{8\pi^{N+1}}
        \vv g\left(\frac{N|z|^2}{2},\vv \right)\,,
\end{equation}
where the function $g(x,y)$ is defined by
\begin{align}\label{jpdf-exact_g}
  g(x,y)  = & \frac{e^{xy/2} }{N(N-1)}
  \left\{x^Ne^{-x} (N-1-x+xy) + \Gamma(N,x)\right.\nonumber\\
  &\times \left.\left[(N-1)(N-2x+xy) + (1-y)x^2 \right] \right\},
\end{align}
with  $x=\frac{N}{2}|z|^2$, $y=\vv$, and the upper incomplete gamma function $\Gamma(a,x)=\int_x^\infty t^{a-1}e^{-t}dt$.

\begin{figure}
  \centering
  \includegraphics[width=0.475\textwidth]{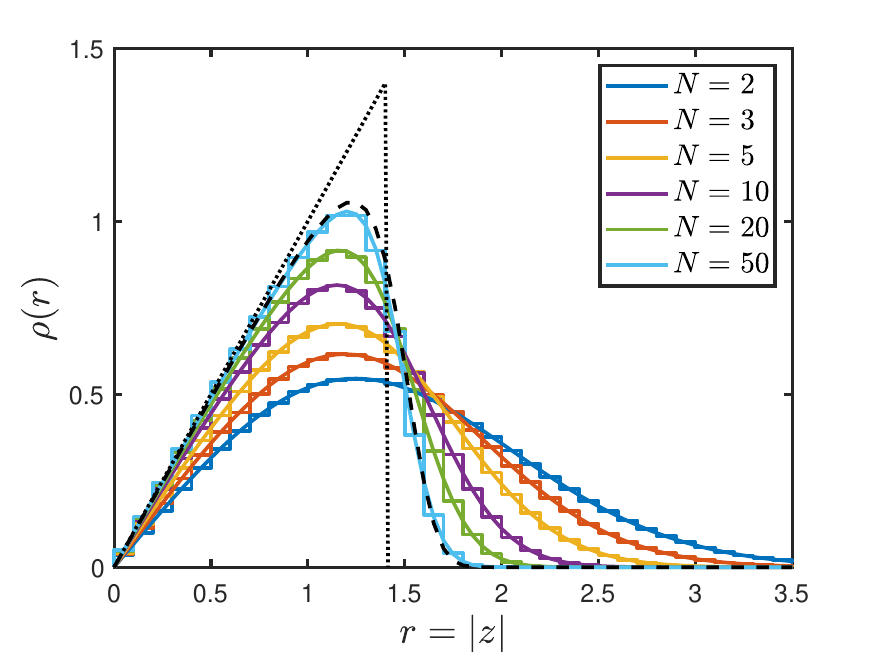}
  \caption{\label{fig:rho}
  The radial spectral density for increasing matrix size $N$. Histograms represent numerical simulations over $10^6$ realizations of Gaussian random matrices for class AI$^{\dagger}$. Solid lines show the exact result (\ref{P(r)}), while the dashed line indicates the large-$N$ approximation (\ref{rho_N}) for $N=50$. The dotted line marks the limiting triangular law for $\rho(r)$ as $N\to\infty$, corresponding to the circular law for $\rho(r)/r$.}
\end{figure}

\textit{Spectral density}.---
As a first application of the above result, we evaluate the eigenvalue density at finite $N$. This density has rotational symmetry and depends only on the radial coordinate $r=|z|$. Hence, it is natural to introduce the radial density $\rho(r) = \pi r \int d^2\vec{v}\mathcal{P}(z,\vec{v})$, normalized as $\int_0^{\infty}\!\rho(r)dr=1$. Integrating the joint distribution (\ref{jpdf-exact}) over $\vec{v}$ can be performed using the following identity, valid for any $N\ge 2$ and any well-behaved function $f$ \cite{fyodorov2025kac}:
\begin{equation}\label{lemma}
  \int d^2\vec{v}\delta(\vec{v}^\dagger\vec{v}-1)f(\vv) = \int_{0}^{1}\frac{dp\,p^{N-2}\pi^N}{\Gamma(N-1)}f(1-p^2)\,.
\end{equation}
Applying this to the radial density leads to the final result
\begin{align}\label{P(r)}
  \rho(r) & = \frac{2\sqrt{N}}{\Gamma(N+2)}\left\{
  \sqrt{\frac{x}{2}}\left[k^2(x) + \left(N-\frac{x}{2}\right)\Gamma(N,x) \right] \right.
  \nonumber \\  & \left.
  + 2^{\frac{N}{2}}\gamma\left(\frac{N+3}{2},\frac{x}{2}\right)
    \left[k(x) + \frac{(N-1-x)\Gamma(N,x)}{2k(x)}\right] \right\},
\end{align}
where we have introduced $k(x)=x^{N/2}e^{-x/2}$, $x=\frac{N}{2}r^2$, and $\gamma(a,x)=\Gamma(a)-\Gamma(a,x)$ for the lower incomplete gamma function. The density behaves as $\rho(r)=\frac{N}{N+1}r+O(r^2)$ near the origin, reaches its maximum around $r\approx\sqrt{2}$ and becomes exponentially suppressed when $r\gg1$, see Fig.~\ref{fig:rho}.

\textit{Nonorthogonality statistics}.--- The distribution $P_r(t)$ of the nonnorthogonality parameter $t$ at a point
with radial coordinate $r$ is determined, up to normalization, by the conditional probability  $\int d^2\vec{v}\mathcal{P}(z,\vec{v})\delta(t+1-\frac{1}{\vv})$. Using the integration identity \eqref{lemma}, one arrives at
\begin{equation}\label{P(t)}
    P_r(t) = \frac{N^2 r}{\Gamma(N+2)\rho(r)}g\left(\frac{Nr^2}{2},\frac{1}{1+t}\right)P_0(t)\,,
\end{equation}
where $g(x,y)$ is as defined in Eq.~(\ref{jpdf-exact_g}) and $P_0(t)$ denotes
\begin{equation}\label{P0(t)}
  P_0(t) = \frac{N^2-1}{4(1+t)^3}\left(\frac{t}{1+t}\right)^{\frac{N-3}{2}}.
\end{equation}
Function $P_0(t)$ has the meaning of the distribution of nonnorthogonality parameter at the spectral origin.
The distribution (\ref{P0(t)}) has the mean $\aver{t}=(N-1)/2$, while all higher moments diverge due to a power-law tail $\sim t^{-3}$. The same tail persists for any value of $r$ (see also Eq.~(\ref{P(tau)}) below).

\begin{figure}
  \centering
  \includegraphics[width=0.475\textwidth]{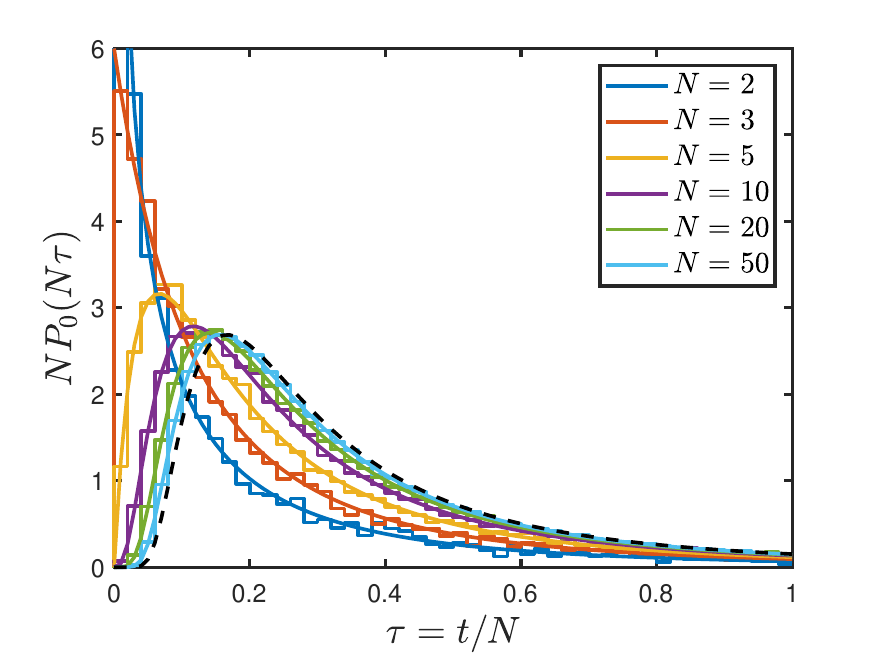}
  \caption{\label{fig:P(t)}
  The distribution of the (rescaled) nonorthogonality factor at the spectral origin for increasing matrix size $N$. Histograms stand for the numerics (the same as in Fig.~\ref{fig:rho}), whereas solid lines correspond to the exact result (\ref{P0(t)}). As $N$ grows, the distribution converges to the limiting form given by Eq.~(\ref{P(tau)}), with $\aver{\tau}=\frac{1}{2}$ at $r=0$, which is shown by the dashed line.}
\end{figure}

All the analytical results obtained above are valid for any matrix size $N\ge 2$. We have verified their validity for different values of $N$ by performing a detailed comparison with the results of Monte-Carlo simulations, shown in
Figs.~\ref{fig:rho} and ~\ref{fig:P(t)}.

\textit{Universal large-$N$ results}.--- The obtained expressions are well suited for the asymptotic analysis in the limit $N\gg1$, when spectral characteristics are expected to become universal. To this end, it is convenient to introduce a scaled variable $s=\sqrt{N}(\frac{x}{N}-1) = \sqrt{N}(\frac{r^2}{2}-1)$. In the $N\to\infty$ limit the point $s=0$ marks the location of the spectrum edge, which in the chosen matrix normalization corresponds to $r=\sqrt{2}$ \cite{radius}. The density behavior near the edge can be established by keeping $s$ fixed as $N\to \infty$, and making use of asymptotic expressions for $k(x)$ and the incomplete gamma-functions, see the End Matter for details. This gives an asymptotic form
\begin{equation}\label{rho_N}
  \rho_{N\gg1}(r)= r\,\Theta_{\mathrm{AI}^\dagger}\left[\sqrt{N}\left(\frac{r^2}{2}-1\right)\right],
\end{equation}
where the edge profile function $\Theta_{\mathrm{AI}^\dagger}(s)$ is given by
\begin{align}\label{edgeprof}
  \Theta_{\mathrm{AI}^\dagger}(s) =&
  \frac{1}{4}\mathrm{erfc}(\frac{s}{\sqrt{2}})
     + \frac{1}{2\sqrt{2}}\mathrm{erfc}\left(-\frac{s}{2}\right)
  \nonumber \\
  &  \times\left[e^{-\frac{s^2}{4}}-\sqrt{\frac{\pi}{2}}\frac{s}{2}e^{\frac{s^2}{4}}
  \mathrm{erfc}\left(\frac{s}{\sqrt{2}}\right)\right]
  \,,
\end{align}
with $\mathrm{erfc}(x)=1-\mathrm{erf}(x)$ being the complimentary error function. Numerically, we found that the obtained expression provides a good approximation already at $N\sim50$ and becomes almost indistinguishable from the exact result for $N\gtrsim100$.

Function $\Theta_{\mathrm{AI}^\dagger}(s)$ exhibits a step-like behavior near $s\sim0$, changing sharply from its limiting value $\Theta_{\mathrm{AI}^\dagger}(s\to-\infty)=1$ to $\Theta_{\mathrm{AI}^\dagger}(s\to+\infty)=0$. In the limit $N\to\infty$, this implies that the radial density follows a `triangular' law, $\rho_\infty(r)=r$ for $r\in[0,\sqrt{2}]$ and vanishes otherwise. Consequently,  the eigenvalues obey the circular law in the complex plane, being uniformly distributed in a disk of radius $\sqrt{2}$. This is a global universal property common to all Ginibre ensembles \cite{byun2025progress}.

We note, however, that the edge profile of the mean density described by Eq.~(\ref{edgeprof}) is distinctly different from one shared  by both complex Ginibre ensemble and the real Ginibre ensembles away from the real axis \cite{byun2025progress}. Such a distinct edge behavior is thus the first clear manifestation of the different universality class of complex symmetric matrices.

The typical nonorthogonality parameter $t$ scales in the large $N$ limit as $t=N\tau$ inside the spectral bulk, with $\tau$ kept fixed. The limiting distribution for $\tau$ takes the following form:
\begin{equation}\label{P(tau)}
  \lim_{N\to\infty} NP_r(N\tau) = \frac{\left\langle \tau\right\rangle^2}{\tau^3} e^{-\frac{\left\langle \tau\right\rangle}{\tau}},
\end{equation}
where $\langle\tau\rangle=\frac{1}{2}(1-\frac{r^2}{2})$ is the mean nonorthogonality factor. We note that although the shape of this limiting distribution is shared with the one for complex Ginibre \cite{bourgadedoubach2020,fyodorov2018CMP}, the value $\left\langle\tau\right\rangle$ at a given point $r$ inside the support is twice as small.

The edge behavior of the distribution $P_r(t)$ requires a different scaling of the nonorthogonality parameter $t=\sqrt{N}\sigma$, with $\sigma$ fixed, keeping $r^2=2(1+\frac{s}{\sqrt{N}})$ as before. We find
\begin{align}\label{P(sig)}
  \lim_{N\to\infty} \sqrt{N}P_r(\sqrt{N}\sigma)
    =& \frac{e^{-\frac{1}{4\sigma^2}+\frac{s}{2\sigma}}}{8\Theta_{\mathrm{AI}^\dagger}(s)\sigma^3}
    \left[ \sqrt{\frac{2}{\pi}}
    \Bigl(\frac{1}{\sigma}-s\Bigr)e^{-\frac{s^2}{2}} \right.\nonumber\\
    &+ \left.(1-\frac{s}{\sigma}+s^2)\,\mathrm{erfc}\Bigl(\frac{s}{\sqrt{2}}\Bigr)\right].
\end{align}
This behavior also differs from that known for complex Ginibre matrices \cite{fyodorov2018CMP} and constitutes another universal feature of the AI$^{\dagger}$ symmetry class. A detailed comparison of the universal edge laws (\ref{edgeprof}) and (\ref{P(sig)}) with their Ginibre counterparts is provided in the End Matter, see Fig.~\ref{fig:A1}.
\begin{figure}
  \centering
  \includegraphics[width=0.49\textwidth]{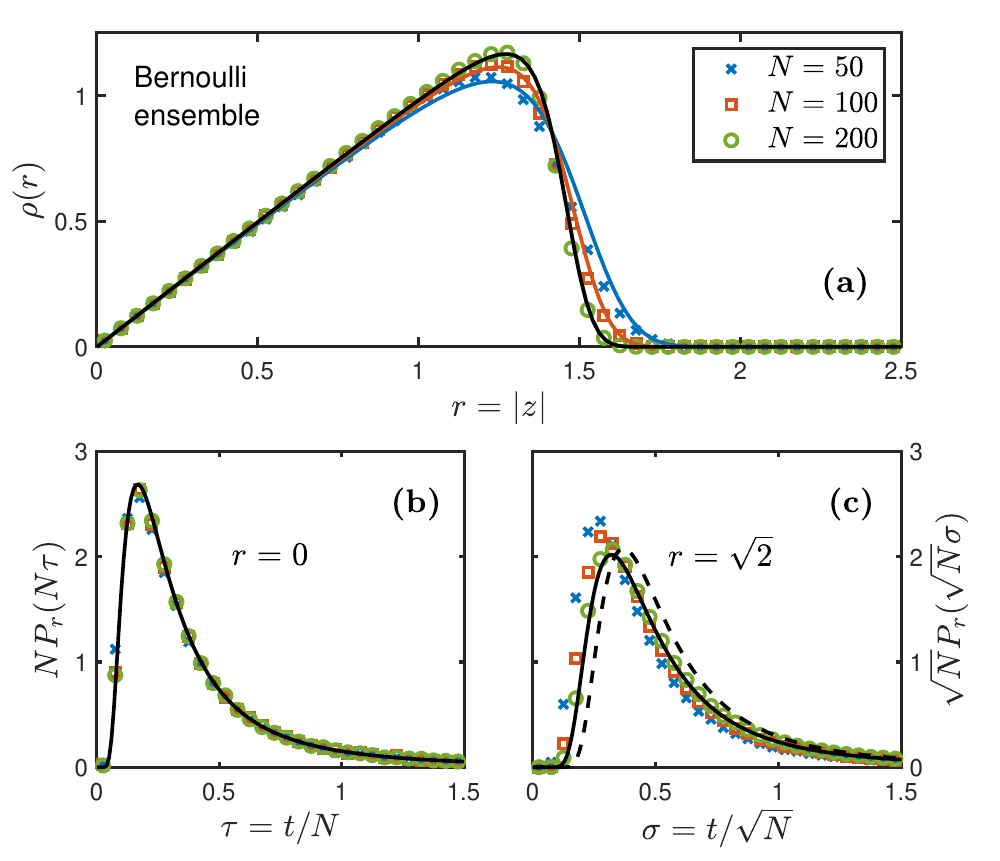}
  \caption{\label{fig:Bern}
  Numerical results for Bernoulli complex symmetric random matrices based on $10^5$ realization of size $N=50\;(\times)$, $100\;(\Box)$ and $200\;(\circ)$. (a) Spectral density $\rho(r)$: the solid lines show our large-$N$ prediction (\ref{rho_N}). (b) Nonorthogonality distribution $P_r(t)$ at the spectral origin ($r=0$): the solid line indicates the bulk limiting form (\ref{P(tau)}). (c) The behavior of $P_r(t)$ at the spectral edge ($r=\sqrt{2}$). The solid line represents our exact expression (\ref{P(t)}) for $N=200$, while the dashed line shows the edge limiting form (\ref{P(sig)}). Note that, due to the  different asymptotic scaling of $t$ with $N$, the convergence to the asymmtotic regime is slower at the edge than in the bulk.
  }
\end{figure}

To assess the universality of our large-$N$ predictions, we performed numerical tests using a non-Gaussian ensemble, choosing complex symmetric random matrices with Bernoulli distributed entries. In this case, each matrix entry (up to transposition symmetry) is an independent complex number whose real and imaginary parts take values $\pm \frac{1}{\sqrt{N}}$ with equal probability. Figure~\ref{fig:Bern}  shows our results for both spectral and nonorthogonality statistics, which demonstrate clear convergence toward the universal regime as $N \gg 1$. This agreement strongly indicates that the asymptotic behavior extends beyond Gaussian ensembles to a broader class of complex symmetric matrices. This is consistent with local bulk universality for nearest-neighbor spacing distributions observed in \cite{hamazaki2020universality}.

\textit{Discussion}.---In summary, we have derived and analyzed the eigenvalue density and the distribution of the diagonal non-orthogonality factor $t=\mathcal{O}_{\mathrm{diag}}-1$ for complex symmetric Gaussian matrices of arbitrary size $N\ge 2$. To address this challenging problem rigorously, we developed a powerful approach that provides simultaneously direct access to eigenvalues and eigenvectors of this class of non-Hermitian random matrices. We expect this framework to be broadly applicable to other non-Hermitian settings. In the large-$N$ limit, when the results serve as universal representatives of the class AI$^\dag$ of quantum chaotic dissipative systems, the global properties in the spectral bulk resemble those of the standard Ginibre matrices. In contrast, the spectral edge statistics exhibit clear and distinctive features specific to this symmetry class, highlighting a fundamental departure from Ginibre universality. The different bulk and edge scaling forms of the non-orthogonality distribution established here should prove important for future applications, e.g., in quantifying universal relaxation behavior in Markovian open quantum systems \cite{bao2025initial}.

We further note that, as argued already by Chalker and Mehlig \cite{ChalkerMehlig2000}, large values of the non-orthogonality factor are controlled by rare events where an eigenvalue $z_k$ comes anomalously close to another eigenvalue. This mechanism was made rigorous for the Ginibre case by the representation of $\mathcal{O}_{\mathrm{diag}}$ in Theorem 2.2 of \cite{bourgadedoubach2020}. Hence the fact that the $t\gg 1$ tail of the non-orthogonality factor distribution for complex symmetric random matrices retains the same cubic form as in complex Ginibre \cite{bourgadedoubach2020,fyodorov2018CMP} for any $N\ge 2$ strongly suggests that eigenvalue repulsion in the bulk is governed primarily by the same factor $|z_i-z_j|^2$ (possibly with logarithmic corrections known to occur for $N=2$ \cite{jaiswal2019universality,hamazaki2020universality}).
Reconciling this observation with the excellent numerical fit of
eigenvalue spacing distributions with $\beta=1.4$ Coulomb gas statistics \cite{akemann2022spacing,akemann2025two} remains an open challenge for theory.

\begin{acknowledgments}
The research at Bielefeld University was supported by the Deutsche Forschungsgemeinschaft (DFG) grant SFB 1283/2 2021–317210226 and at King's College London  by  EPSRC grant UKRI1015 "Non-Hermitian random matrices: theory and applications". YVF is grateful for the hospitality of the program "Lush World of Random Matrices" at ZiF, Bielefeld University, during the final stages of this project.
\end{acknowledgments}
%

\section{End Matter}
\setcounter{secnumdepth}{2}
\setcounter{equation}{0}
\renewcommand{\theequation}{A\arabic{equation}}
\setcounter{figure}{0}
\renewcommand{\thefigure}{A\arabic{figure}}
\subsection*{Appendix: Large $N$ asymptotics}
To establish the large-$N$ behavior of the spectral density, we require the leading-order asymptotic forms of the function $k(x)=x^{N/2}e^{-x/2}$ and of the incomplete lower and upper gamma-functions appearing in Eq.~(\ref{P(r)}) at $N\to \infty$. These forms depend on the spectral parameter $r$. Accordingly, two different regimes arise: the global scaling limit in the bulk, where $x=Np$ and $p\in[0,1)$ is fixed, and the local scaling limit at the edge, where $p=1+\frac{s}{\sqrt{N}}$ with $s$ fixed. Using standard asymptotic analysis \cite{NIST}, we obtain the required form 
\begin{equation}
\frac{k(Np)}{N^{N/2}} \approx\left\{
\begin{array}{cl}
  p^{\frac{N}{2}}e^{-Np/2},  &\quad\mbox{bulk}\\
  e^{-N/2}e^{-s^2/4},        &\quad\mbox{edge}
\end{array}\right..
\end{equation}
The corresponding asymptotic forms for the incomplete lower and upper gamma-functions are respectively given by
\begin{equation}\label{gamma}
\frac{\gamma\left(\frac{N+3}{2}, \frac{Np}{2}\right)}{(N/2)^{\frac{N+1}{2}}} \approx\left\{
\begin{array}{cl}
  p^{\frac{N+3}{2}}e^{-Np/2}\frac{1}{1-p},
  &\quad\mbox{bulk}\\
  e^{-N/2}\sqrt{\frac{\pi}{2}}\mbox{erfc}{\left(-\frac{s}{2}\right)},
  &\quad\mbox{edge}
\end{array}\right.
\end{equation}
and
\begin{equation}\label{Gamma}
\frac{\Gamma\left(N, Np\right)}{\Gamma(N)}\approx \left\{
\begin{array}{cl}
  1, &\quad\mbox{bulk}\\
  \frac{1}{2}\mbox{erfc}{\left(\frac{s}{\sqrt{2}}\right)}, &\quad\mbox{edge}
\end{array}\right..
\end{equation}

For the distribution $P_r(t)$ given by Eq.~(\ref{P(t)}), the nonorthogonality factor $t$ exhibits different scaling with $N$: in the bulk we have $t=N\tau$, with $\tau$ fixed, whereas at the spectral edge $t=\sqrt{N}\sigma$, with $\sigma$ fixed. Correspondingly, the function $P_0(t)$ from Eq.~(\ref{P0(t)}) takes the following asymptotic form:
\begin{equation}
 P_0(t) \approx \left\{
\begin{array}{cl}
  \frac{1}{4N\tau^3}e^{-\frac{1}{2\tau}}, &\quad\mbox{bulk}\\
  \frac{\sqrt{N}}{4\sigma^3}e^{-\frac{\sqrt{N}}{2\sigma}+\frac{1}{4\sigma^2}}, &\quad\mbox{edge}
\end{array}\right..
\end{equation}
Finally, for the function $e^{xy/2}$ with $y=\frac{1}{1+t}$, we find
\begin{equation}
 e^{xy/2} \approx \left\{
\begin{array}{cl}
  e^{-\frac{p}{2\tau}}, &\quad\mbox{bulk}\\
  e^{\frac{\sqrt{N}}{2\sigma}+\frac{s}{2\sigma}-\frac{1}{2\sigma^2} }, &\quad\mbox{edge}
\end{array}\right..
\end{equation}
Note that the convergence to the asymmtotic regime is slower at the edge than in the bulk, owing to the different scaling of $t$ with $N$, see also Fig.~\ref{fig:Bern}(c). The accuracy of the asymptotic edge result can be improved by using the exact representation
\begin{equation}
  e^{xy/2} = \exp\left[\frac{1}{2\sigma}\left(\sqrt{N}+
  \frac{s -\frac{1}{\sigma}}{1+\frac{1}{\sqrt{N}\sigma}}\right)\right]\,,
\end{equation}
retaining more next-to-leading terms in the $\frac{1}{\sqrt{N}}$-expansion.

\begin{figure}
  \centering
  \includegraphics[width=0.475\textwidth]{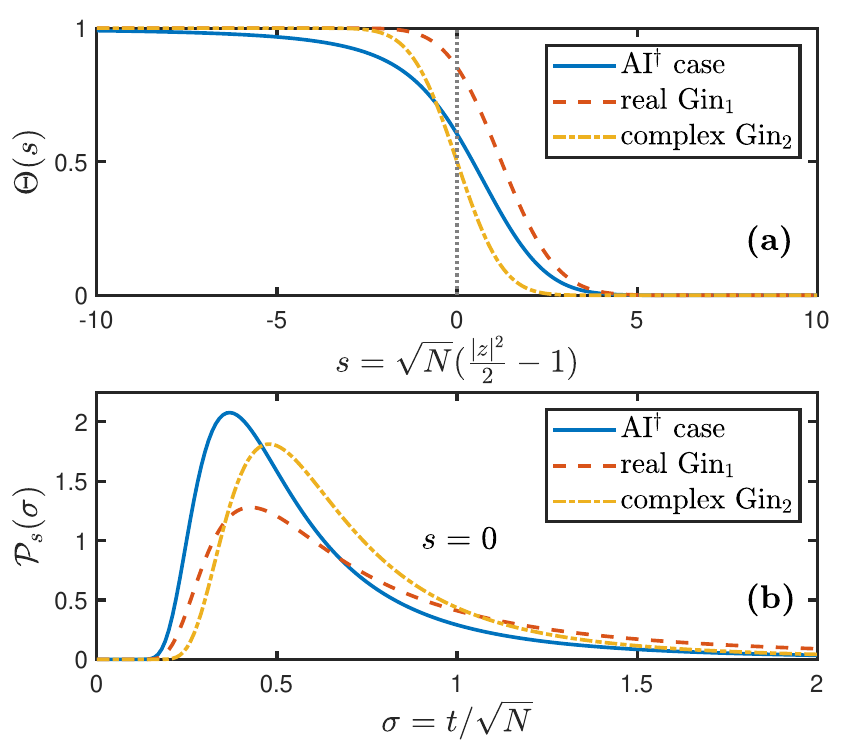}
  \caption{\label{fig:A1}
  Large-$N$ asymptotic forms for the complex-symmetric (this work) and the two Ginibre ensembles: (a) the spectral edge profile; (b) the nonorthogonality distribution at the edge ($s=0$).}
\end{figure}

Finally, it is instructive to compare the edge behavior obtained here for the complex-symmetric case with that known for the real and complex Ginibre ensembles \cite{byun2025progress}. Using the same normalization for the spectral edge at $r=\sqrt{2}$, the edge profile for complex Ginibre (denoted by Gin$_2$) is given by
\begin{equation}\label{theta2}
  \Theta_2(s) =\frac{1}{2}\mbox{erfc}\left(\frac{s}{\sqrt{2}}\right),
\end{equation}
with fixed $s=\sqrt{N}(\frac{|z|^2}{2}-1)$ at any point $z$ in the complex plane. In the case of real Ginibre (denoted by Gin$_1$), one must distinguish between complex eigenvalues and those lying on the real axis, as they exhibit different statistical properties in the $N\to\infty$ limit. The latter have the following edge profile:
\begin{equation}\label{theta1}
  \Theta_1(s) = \frac{1}{2}\left[\mbox{erfc}\left(\frac{s}{\sqrt{2}}\right)
  +\frac{1+\mbox{erfc}\left(\frac{s}{2}\right)}{\sqrt{2}} e^{-\frac{s^2}{4}}\right]\,,
\end{equation}
where now $s$ refers to the real eigenvalues, whereas the complex eigenvalues share the same edge statistics as in Gin$_2$.

The edge profiles exhibit a distinctly different behavior only within a narrow region near the edge ($s=0$), while for negative $s$ they rapidly approach the bulk limit of unity, see Fig.~\ref{fig:A1}. This accounts for the high accuracy of the large-$N$ asymptotic form (\ref{rho_N}), even for moderate values of $N$.

The edge statistics of the nonorthogonality factor is also known in these two Ginibre cases
\cite{fyodorov2018CMP}. The limiting distribution $\mathcal{P}_s(\sigma)=\lim_{N\to\infty}\sqrt{N}P_r(\sqrt{N}\sigma)$ for real Gin$_1$ reads
\begin{align}\label{p_1(sigma)}
  \mathcal{P}_{s,1}(\sigma)     = \frac{e^{-\frac{1}{4\sigma^2}+\frac{s}{2\sigma}}}{2\Theta_1(s)\sigma^2}
    \left[\frac{e^{-\frac{s^2}{2}}}{\sqrt{2\pi}}
    + \frac{1-s\sigma}{2\sigma}\mathrm{erfc}\Bigl(\frac{s}{\sqrt{2}}\Bigr)\right].
\end{align}
For the Gin$_2$ case, the distribution is given by
\begin{align}\label{p_2(sigma)}
  \mathcal{P}_{s,2}(\sigma) =& \frac{e^{-\frac{1}{2\sigma^2}+\frac{s}{\sigma}}}{2\Theta_2(s)\sigma^3}
    \left[\Bigl(2+\frac{s}{\sigma}-\frac{1}{\sigma^2}\Bigr)\frac{e^{-s^2}}{\pi}
    \right. \nonumber\\
    & - \Bigl(3s-\frac{1-s^2}{\sigma}-\frac{s}{\sigma^2}\Bigr)
    \frac{e^{-\frac{s^2}{2}}}{\sqrt{2\pi}}
         \mathrm{erfc}\Bigl(\frac{s}{\sqrt{2}}\Bigr)
    \nonumber\\
    & + \left.\frac{1}{2}\Bigl[\Bigl(s-\frac{1}{\sigma}\Bigr)^2-1\Bigr]
        \mathrm{erfc}^2\Bigl(\frac{s}{\sqrt{2}}\Bigr)\right].
\end{align}

Figure~\ref{fig:A1} illustrates these edge profiles and nonorthogonality distributions, clearly revealing a distinct universal scaling behavior for the three ensembles discussed above. We expect this universality to persist beyond the Gaussian case, i.e. the scaling forms to depend only on the underlying symmetry class, rather than on the specific distribution of matrix entries.

\clearpage
\widetext
\setcounter{secnumdepth}{2}
\begin{center}
 \textbf{\large Supplemental Material for \\ \smallskip ``\MyTitle''}
\\ \smallskip by Gernot Akemann, Yan V. Fyodorov, and Dmitry V. Savin
\end{center}
\setcounter{equation}{0}
\renewcommand{\theequation}{S\arabic{equation}}
\setcounter{figure}{0}
\renewcommand{\thefigure}{S\arabic{figure}}
\setcounter{table}{0}
\renewcommand{\thetable}{S\arabic{table}}
\setcounter{section}{0}
\renewcommand{\thesection}{S-\Roman{section}}
This Supplemental Material provides additional details of the calculations outlined in the main text. The first section describes the approach used to derive the integral representation (\ref{jpdf-int}). The following section develops an efficient integration method based on Wick’s theorem, leading to the main final result (\ref{jpdf-exact}).

\section{Supersymmetry method and integral representation}
\label{SM1}
\subsection{Ensemble averages}
We will use the simplified notations $\vec{a}^T\vec{b} = (\vec{a}\cdot\vec{b})$ and $\vec{a}\vec{b}^T= \vec{a}\otimes\vec{b}$  to denote the inner and outer products, respectively.
It is convenient to start with the form $J=X+iY$ and represent the eigenvalue problem for a complex eigenvalue $z=\lambda_1+i\lambda_2$ with the right eigenvector $\vec{v}=\vec{s}+i\vec{t}$ in terms of the relevant real ($X,\lambda_1,\vec{s}$) and imaginary ($Y,\lambda_2,\vec{t}$) parts. The eigenvector is normalized according to $\vec{v}^\dagger\vec{v}=\vec{s}^T\vec{s}+\vec{t}^T\vec{t}=1$. Our starting point is the formula (\ref{jpdf-def}) which can then be expressed as
\begin{equation}\label{SM:jpdf-real}
  \mathcal{P}(z,\vec{v}) = \frac{\delta(\vec{v}^\dagger\vec{v}-1)}{\pi N}
  \aver{
   \delta\left[(X-\lambda_1)\vec{s}-(Y-\lambda_2)\vec{t}\right]
   \delta\left[(Y-\lambda_2)\vec{s}+(X-\lambda_1)\vec{t}\right]
   \left|\frac{d}{dz}\det(J-z)\right|^2 },
\end{equation}
involving only the delta-functions of a real-valued vector argument. We can now employ the Fourier representation
\begin{equation}\label{SM:delta}
  \delta(\vec{x}_{\sigma}) \equiv \prod_{i=1}^{N}\delta(x_{\sigma,i})
  = \int \frac{d\vec{k}_\sigma}{(2\pi)^N} \exp(i\vec{k}_\sigma^T\vec{a}_\sigma)\,,
  \qquad d\vec{k}_\sigma=\prod_{i=1}^{N}dk_{\sigma,i}\,, \quad \sigma=1,2\,,
\end{equation}
for each of the two vector delta-functions in the product $\delta(\vec{x}_1)\delta(\vec{x}_2)$ above and cast the JPDF as follows:
\begin{equation}\label{SM:jpdf-real_b}
  \mathcal{P}(z,\vec{v}) = \frac{\delta(\vec{v}^\dagger\vec{v}-1)}{\pi N}
  \iint \frac{d\vec{k}_1d\vec{k}_2}{(2\pi)^{2N}}
   e^{-i\lambda_1(\vec{k}_1^T\vec{s}+\vec{k}_2^T\vec{t}) + i\lambda_2(\vec{k}_1^T\vec{t}-\vec{k}_2^T\vec{s})}
  \aver{ e^{i\Tr[X(\vec{s}\vec{k}_1^T+\vec{t}\vec{k}_2^T)]
          + i\Tr[Y(\vec{s}\vec{k}_2^T-\vec{t}\vec{k}_1^T)] }
   \left|\frac{d}{dz}\det(J-z)\right|^2 }.
\end{equation}
Here, we have used the identity $f(\vec{a}^T\vec{b})=f(\Tr [\vec{b}\vec{a}^T])$  which follows from the invariance of a trace under cyclic permutations.

Next, we can use a Hermitization trick
$ \left|\frac{d}{dz}\det(J-z)\right|^2 = (-1)^N \frac{\partial^2}{\partial{z}\,\partial{z^*}}
  \det\begin{pmatrix} 0 & J-z \\ J^\dagger-z^* & 0 \end{pmatrix}$
by doubling the matrix dimension, and then represent the determinant as the Gaussian
integral of Berezin type
\begin{equation}\label{SM:det}
  \det\begin{pmatrix} 0 & J-z \\ J^\dagger-z^* & 0 \end{pmatrix}
  = (-1)^N \int D(\vec{\psi},\vec{\phi})
  \exp\left[ -(\vec{\psi}_1^T,\vec{\psi}_2^T)
    \begin{pmatrix} 0 & J-z \\ J^\dagger-z^* & 0 \end{pmatrix}
  \begin{pmatrix} \vec{\phi}_1 \\ \vec{\phi}_2\end{pmatrix} \right],
\end{equation}
where the vectors $\vec{\psi}_\sigma=(\psi_{\sigma,1},\ldots,\psi_{\sigma,N})^T$ and $\vec{\phi}_\sigma=(\phi_{\sigma,1},\ldots,\phi_{\sigma,N})^T$, with $\sigma=1,2$, contain anticommuning (Grassmann) variables only, and the integration measure $D(\vec{\psi},\vec{\phi}) = \prod_{\sigma=1,2}\prod_{i=1}^{N}d\psi_{\sigma,i}d\phi_{\sigma,i}$. All these steps combined, the result is
\begin{align}\label{SM:jpdf-real_bf}
  \mathcal{P}(z,\vec{v}) = \frac{\delta(\vec{v}^\dagger\vec{v}-1)}{\pi N}
  \!\!\iint\!\frac{d\vec{k}_1d\vec{k}_2}{(2\pi)^{2N}}
   e^{-i\lambda_1(\vec{k}_1^T\vec{s}+\vec{k}_2^T\vec{t}) + i\lambda_2(\vec{k}_1^T\vec{t}-\vec{k}_2^T\vec{s})}
    \frac{\partial^2}{\partial{z}\,\partial{z^*}}\!\!\int\!D(\vec{\psi},\vec{\phi})
     e^{z(\vec{\psi}_1^T\vec{\phi}_2) + z^*(\vec{\psi}_2^T\vec{\phi}_1)}
     \aver{e^{i\Tr(XT_x) + i\Tr(YT_y)}},
\end{align}
where we have introduced the following two matrices:
\begin{equation}\label{SM:T_xy}
  T_x = (\vec{s}\vec{k}_1^T + \vec{t}\vec{k}_2^T)
       - i(\vec{\phi}_2\vec{\psi}_1^T + \vec{\phi}_1\vec{\psi}_2^T)
       \quad\mbox{and}\quad
  T_y = (\vec{s}\vec{k}_2^T - \vec{t}\vec{k}_1^T)
       + (\vec{\phi}_2\vec{\psi}_1^T - \vec{\phi}_1\vec{\psi}_2^T)\,.
\end{equation}

The obtained expression (\ref{SM:jpdf-real_bf}) is well suited for statistical averaging. Since $X$ and $Y$ are statistically independent, the last term $\aver{\cdots}$ in Eq.~(\ref{SM:jpdf-real_bf}) factorizes into the product $\aver{e^{i\Tr(XT_x)}}_X\aver{e^{i\Tr(YT_y)}}_Y$, representing two separate averages over GOE matrices. Due to the symmetry $X^T=X$, the GOE averaging requires symmetrization and can be carried over as follows:
\begin{equation}\label{SM:aver_X}
  \aver{e^{i\Tr(XT_x)}}_X = \aver{e^{\frac{i}{2}\Tr[X(T_x+T_x^T)]}}_X
  = e^{-\frac{1}{4N}\Tr(T_x+T_x^T)^2}
\end{equation}
and similarly for $Y$. To this end, it is convenient to specify the symmetrized versions of the matrices $T_{x,y}$ and introduce
\begin{equation}\label{SM:T_xy_sym}
 \begin{array}{l}
  T_x+T_x^T =
  (\vec{s}\vec{k}_1^T + \vec{k}_1\vec{s}^T + \vec{t}\vec{k}_2^T + \vec{k}_2\vec{t}^T)
  - i(\vec{\phi}_2\vec{\psi}_1^T - \vec{\psi}_1\vec{\phi}_2^T + \vec{\phi}_1\vec{\psi}_2^T-\vec{\psi}_2\vec{\phi}_1^T)
  \equiv B_+ - iF_+\\[1ex]
  T_y+T_y^T = (\vec{s}\vec{k}_2^T + \vec{k}_2\vec{s}^T - \vec{t}\vec{k}_1^T-\vec{k}_1\vec{t}^T)
       + (\vec{\phi}_2\vec{\psi}_1^T -\vec{\psi}_1\vec{\phi}_2^T- \vec{\phi}_1\vec{\psi}_2^T+\vec{\psi}_2\vec{\phi}_1^T)
  \equiv B_- + F_-
 \end{array} \,.
\end{equation}
Note that these matrices contain both ordinary (``bosonic'', $B_\pm$) and nilpotent (``fermionic'', $F_\pm$) parts arising from anticommuting variables. In particular, this leads to the useful identity $\Tr(F_\pm)^2=\mp4[(\vec{\psi}_1^T\vec{\phi}_1)(\vec{\psi}_2^T\vec{\phi}_2)- (\vec{\phi}_2^T\vec{\phi}_1)(\vec{\psi}_2^T\vec{\psi}_1)]$.
Substituting these forms into (\ref{SM:aver_X}), we arrive after straightforward but lengthy algebra at the average factor expressed as
\begin{align}\label{SM:aver}
 \aver{e^{i\Tr(XT_x) + i\Tr(YT_y)}} = I_{BB}I_{BF}I_{FF}\,,
\end{align}
which in turn leads to the following integral representation of the JPDF:
\begin{align}\label{SM:jpdf-susy}
  \mathcal{P}(z,\vec{v}) = \frac{\delta(\vec{v}^\dagger\vec{v}-1)}{\pi N}
  \!\!\iint\!\frac{d\vec{k}_1d\vec{k}_2}{(2\pi)^{2N}}
  e^{-i\lambda_1(\vec{k}_1^T\vec{s}+\vec{k}_2^T\vec{t}) + i\lambda_2(\vec{k}_1^T\vec{t}-\vec{k}_2^T\vec{s})} I_{BB}
  \frac{\partial^2}{\partial{z}\,\partial{z^*}}\!\!\int\!D(\vec{\psi},\vec{\phi})
  e^{z(\vec{\psi}_1^T\vec{\phi}_2) + z^*(\vec{\psi}_2^T\vec{\phi}_1)} I_{BF} I_{FF},
\end{align}
where
\begin{subequations}
\begin{align}
\label{SM:I_BB}
& I_{BB} = e^{-\frac{1}{2N}[ \vec{k}_1^T\vec{k}_1 + \vec{k}_2^T\vec{k}_2
  + (\vec{k}_1^T\vec{s})^2 + (\vec{k}_1^T\vec{t})^2 + (\vec{k}_2^T\vec{s})^2 + (\vec{k}_2^T\vec{t})^2 + 2(\vec{k}_1^T\vec{t})(\vec{k}_2^T\vec{s})^2 - 2(\vec{k}_1^T\vec{s})(\vec{k}_2^T\vec{t})^2 ]}\,,  \\
\label{SM:I_BF}
& I_{BF} = e^{-\frac{1}{2N}[ \vec{\psi}_1^T(iB_+-B_-)\vec{\phi}_2
  + \vec{\phi}_2^T(-iB_++B_-)\vec{\psi}_1 + \vec{\psi}_2^T(iB_++B_-)\vec{\phi}_1
  + \vec{\phi}_1^T(iB_++B_-)\vec{\psi}_2 ]}\,,  \\
\label{SM:I_FF}
& I_{FF} = e^{-\frac{2}{N}[ (\vec{\psi}_1^T\vec{\phi}_1)(\vec{\psi}_2^T\vec{\phi}_2)
  - (\vec{\phi}_2^T\vec{\phi}_1)(\vec{\psi}_2^T\vec{\psi}_1) ]}\,.
\end{align}
\end{subequations}

As is typical for the supersymmetry approach, the resulting expression is bi-quadratic in the integration vectors. Conventional techniques such as the Hubbard–Stratonovich transformation or bosonization are usually employed to perform the integration by reducing it to one over a certain supermatrix. However, we found those methods to be less efficient here due to the algebraic complexity of the expressions involved. To overcome this difficulty, we adopt a different strategy: we first carry out the exact integration over the anticommuting variables. This reduces the problem to an integration involving a Pfaffian, as described next.

\subsection{Reduction to a Pfaffian problem}
To integrate over the Grassmann variables, we first need to decouple the bi-quadratic terms appearing in the factor $I_{FF}$ of Eq.~(\ref{SM:aver}). This can be accomplished using the following identities, valid for any real $a,b$:
\begin{equation}\label{SM:decouple}
  e^{\frac{1}{N}ab} = \frac{N}{\pi}\int d^2q e^{-N|q|^2-qa-q^*b}
  \quad\mbox{and}\quad
  e^{-\frac{1}{N}ab}= \frac{N}{\pi}\int d^2q e^{-N|q|^2-iqa-iq^*b}\,,
  \qquad d^2q=d(\Re{q})\,d(\Im{q})\,.
\end{equation}
Introducing auxiliary integrations over two complex variables $q_{1,2}$, we obtain
\begin{equation}\label{SM:I_FF_int}
  I_{FF} = \left(\frac{N}{2\pi}\right)^2\iint d^2q_1d^2q_2 \,
  e^{ -\frac{N}{2}(|q_1|^2+|q_2|^2)
   -q_1(\vec{\phi}_2^T\vec{\phi}_1) - q_1^*(\vec{\psi}_2^T\vec{\psi}_1)
   -iq_2(\vec{\psi}_1^T\vec{\phi}_1) - iq_2^*(\vec{\psi}_2^T\vec{\phi}_2) }.
\end{equation}
Collecting all anticommuting variables into one $4N$-dimensional vector
$\vec{\Psi}^T=(\vec{\psi}_1^T,\vec{\phi}_2^T,\vec{\psi}_2^T,\vec{\phi}_1^T)$, we can further represent
\begin{equation}
  I_{FF} = \left(\frac{N}{2\pi}\right)^2\int d^2\vec{q}\,e^{-\frac{N}{2}|\vec{q}|^2}
   \exp\left[-\frac{1}{2}\vec{\Psi}^T
    \begin{pmatrix} 0 & -Q^T \\
                     Q & 0 \end{pmatrix} \vec{\Psi}\right]\,,
  \qquad  Q \equiv \begin{pmatrix} q_1^* & iq_2^* \\
                      -iq_2 & -q_1 \end{pmatrix} \otimes\id_N\,,
\end{equation}
where we have also combined $q_{1,2}$ into the complex 2-vector $\vec{q}=(q_1,q_2)$, with $d^2\vec{q}=d^2q_1d^2q_2$. Similarly, one can write
\begin{equation}\label{SM:I_BF_int}
  e^{z(\vec{\psi}_1^T\vec{\phi}_2) + z^*(\vec{\psi}_2^T\vec{\phi}_1)} I_{BF}
  = \exp\left[-\frac{1}{2}\vec{\Psi}^T
    \begin{pmatrix} P_-\otimes\tau_2 & 0 \\
                     0 & P_+\otimes\tau_2 \end{pmatrix} \vec{\Psi}\right]\,,
  \qquad
  \begin{array}{l}
    P_+ \equiv z^*\id_N -\frac{1}{N}(B_- + iB_+)  \\[1ex]
    P_- \equiv z \id_N - \frac{1}{N}(B_- - iB_+)
  \end{array}\,,
\end{equation}
where the Pauli matrix $\tau_2=\begin{pmatrix} 0 & -1\\ 1 & 0 \end{pmatrix}$. As a result, the action associated with the anticommuting integration variables in (\ref{SM:jpdf-susy}) become quadratic, $e^{-\frac{1}{2}\vec{\Psi}^T S \vec{\Psi}}$, where the $4N\times4N$ skew-symmetric matrix $S$ is defined by
\begin{equation}
  S =
  \begin{pmatrix} P_-\otimes\tau_2 & -Q^T \\
                  Q & P_+\otimes\tau_2 \end{pmatrix}.
\end{equation}
The resulting Gaussian integral over $\vec{\Psi}$ can then be evaluated explicitly, yielding the Pfaffian of $S$ and leading to
\begin{equation}
  \int\!D(\vec{\psi},\vec{\phi})
  e^{z(\vec{\psi}_1^T\vec{\phi}_2) + z^*(\vec{\psi}_2^T\vec{\phi}_1)} I_{BF}I_{FF}
  = \left(\frac{N}{2\pi}\right)^2\int d^2\vec{q}\,e^{-\frac{N}{2}|\vec{q}|^2}
  \mathrm{Pf}(S)\,.
\end{equation}

The Pfaffian of a block matrix can be computed using the identity $\mathrm{Pf}\begin{pmatrix}A&-Q^T \\ Q & D\end{pmatrix} = \sqrt{\det(DA+DQ^TD^{-1}Q)}$. In our case, we find $DA= -(P_+P_-)\otimes\id_2$ and $DQ^TD^{-1}Q=-(|q_1|^2+|q_2|^2)\id_N\otimes\id_2$, which together yield a block-diagonal structure for the $2N\times2N$ matrix $DA+DQ^TD^{-1}Q=-(|\vec{q}|^2\id_N+P_+P_-)\otimes\id_2$. This structure eliminates the square root and leads to a compact expression for the Pfaffian as the following determinant:
\begin{equation}\label{SM:pfaff-det}
    \mathrm{Pf}(S) = \det(q^2\id_N+P_+P_-),
    \qquad q\equiv|\vec{q}|=\sqrt{|q_1|^2+|q_2|^2}\,.
\end{equation}

This representation significantly simplifies the problem. However, computing the resulting determinant directly making use of algebraic methods proves challenging due to its structure. To address this, we adopt an alternative approach and compute it by introducing an auxiliary Grassmann integration, as detailed below.

\subsection{Computation of the determinant}

We first observe that the algebraic structure of the determinant in (\ref{SM:pfaff-det}) allows it to be represented as a Gaussian integral
\begin{equation}\label{SM:D_N}
  D_N\equiv\det(q^2\id_N+P_+P_-) = \det
  \begin{pmatrix} q\id_N & P_+\\ -P_- & q\id_N\end{pmatrix}
  = \int d[\vec{\chi}] \exp\left[-(\vec{\chi}_1^\dagger,\vec{\chi}_2^\dagger,)
    \begin{pmatrix} q\id_N & P_+\\ -P_- & q\id_N\end{pmatrix}
    \begin{pmatrix} \vec{\chi}_1\\ \vec{\chi}_2\end{pmatrix}
    \right]
\end{equation}
over $2N$-vector $\vec{\chi}^T=(\vec{\chi}_1^T,\vec{\chi}_2^T)$, with complex Grassmann entries $\{\chi_{\sigma,i}\}$, where $d[\vec{\chi}]=\prod_{\sigma=1,2}\prod_{i=1}^{N}d\chi_{\sigma,i}^*d\chi_{\sigma,i}$.
It is convenient at this stage to introduce back the complex vector $\vec{s}+i\vec{t}=\vec{v}$ and define a new complex vector $\vec{k}_2+i\vec{k}_1\equiv N\vec{k}$ (note the scaling with $N$). In terms of these variables, the matrices $B_-\pm iB_+$ from Eq.~(\ref{SM:I_BF_int}) take the following compact form:
\begin{equation}
  B_- + iB_+ = N(\vec{v}\vec{k}^T + \vec{k}\vec{v}^T) \quad\mbox{and}\quad
  B_- - iB_+ = (B_- + iB_+)^* = N(\vec{v}^*\vec{k}^\dagger + \vec{k}^*\vec{v}^\dagger)\,.
\end{equation}
Expanding the quadratic form in Eq.~(\ref{SM:D_N}), we obtain
\begin{equation}\label{SM:D_N_int}
 D_N = \int d[\vec{\chi}] e^{
  -q(\vec{\chi}_1^\dagger\vec{\chi}_1+\vec{\chi}_2^\dagger\vec{\chi}_2)
  -z^*(\vec{\chi}_1^\dagger\vec{\chi}_2)+z(\vec{\chi}_2^\dagger\vec{\chi}_1)
 }
 e^{ (\vec{\chi}_1^\dagger\vec{v})(\vec{k}^T\vec{\chi}_2)
    +(\vec{\chi}_1^\dagger\vec{k})(\vec{v}^T\vec{\chi}_2)
    +(\vec{\chi}_2^\dagger\vec{v}^*)(\vec{k}^\dagger\vec{\chi}_1)
    +(\vec{\chi}_2^\dagger\vec{k}^*)(\vec{v}^\dagger\vec{\chi}_1)
 }.
\end{equation}

Integrating over $\vec{\chi}_{1,2}$ expresses the determinant in terms of all the invariant combinations of the vectors $\vec{v}$ and $\vec{k}$ and their complex conjugates. To perform this, we need to decouple the terms appearing in the last factor above. This can be achieved by introducing an additional Grassmann integration using the following identity, valid for any anticommuting variables $\alpha,\beta$:
\begin{equation}\label{SM:decouple_F}
  \int d\rho^*d\rho\,e^{-\rho^*\rho-\rho^*\alpha-\rho\beta}
  = e^{\alpha\beta} = e^{-\beta\alpha}\,.
\end{equation}
This can be viewed as an anticommuting extension of the standard decoupling identity for commuting variables, Eq.~(\ref{SM:decouple}).

Introducing four additional Grassmann variables, organised as 4-vector $\vec{\rho}=(\rho_1,\rho_2,\rho_3,\rho_4)^T$, we can cast Eq.~(\ref{SM:D_N_int}) as
\begin{equation}\label{SM:D_N_int2}
  D_N  = \int d[\vec{\rho}]\,e^{-\vec{\rho}^\dagger\vec{\rho}}
  \int d[\vec{\chi}]\,e^{-\vec{\chi}^\dagger F \vec{\chi}
  +\vec{\chi}^\dagger\vec{\alpha} -\vec{\beta}^T\vec{\chi} },
\end{equation}
where
\begin{equation}\label{SM:F,a,b}
  F = \begin{pmatrix} q & z^*\\ -z & q\end{pmatrix}\otimes\id_N\,, \qquad
  \vec{\alpha}=\begin{pmatrix} \rho_1^*\vec{v}+\rho_2^*\vec{k}\\
      \rho_3^*\vec{v}^*+\rho_4^*\vec{k}^* \end{pmatrix} \quad\mbox{and}\quad
  \vec{\beta}=\begin{pmatrix} \rho_3\vec{k}^*+\rho_4\vec{v}^*\\
      \rho_1\vec{k}+\rho_2\vec{v} \end{pmatrix}.
\end{equation}
The integration over $\vec{\chi}$ can now be performed using the following identity, valid for any invertable matrix $F$:
\begin{equation}
 \int d[\vec{\chi}]\,e^{-\vec{\chi}^\dagger F \vec{\chi}
  +\vec{\chi}^\dagger\vec{\alpha} -\vec{\beta}^T\vec{\chi}}
 = \det(F)\,e^{-\vec{\beta}^T F^{-1}\vec{\alpha}}.
\end{equation}
This formula provides a vector generalization of the decoupling rule (\ref{SM:decouple_F}).

Using the explicit forms given in Eq.~(\ref{SM:F,a,b}), we first find  $\det{F}=(q^2+|z|^2)^N$ and
$F^{-1}=\frac{1}{q^2+|z|^2}\begin{pmatrix} q & -z^*\\ z & q\end{pmatrix}\otimes\id_N$. Next, evaluating $\vec{\beta}^T F^{-1}\vec{\alpha}$ and performing some algebra, we obtain for $D_N$ from Eq.~(\ref{SM:D_N_int2}) the following result:
\begin{equation}\label{}
D_N  = (q^2+|z|^2)^N\int
  d[\vec{\rho}]\,e^{-\vec{\rho}^\dagger\vec{\rho}-\vec{\rho}^\dagger M\vec{\rho}}
  = (q^2+|z|^2)^N \det(\id_4-M)\,,
\end{equation}
where the $4\times4$ matrix $M$ is defined as follows
\begin{equation}\label{SM:M}
  M = \frac{1}{q^2+|z|^2}\begin{pmatrix}zA & qB\\ qB^* & -z^*A^*\end{pmatrix},
  \qquad\mbox{with}\quad
  A = \begin{pmatrix} \vec{k}^T\vec{v} & \vec{v}^T\vec{v} \\
        \vec{k}^T\vec{k} & \vec{v}^T\vec{k} \end{pmatrix}
  \quad\mbox{and}\quad
  B = \begin{pmatrix} \vec{k}^\dagger\vec{v} & \vec{v}^\dagger\vec{v} \\
        \vec{k}^\dagger\vec{k} & \vec{v}^\dagger\vec{k} \end{pmatrix}\,.
\end{equation}
Since $\vec{v}^\dagger\vec{v}=1$ due to the normalization, we readily find $D_N=(q^2+|z|^2)^{N-4}D_q(z,\vec{v,\vec{k}})$, as stated in Eq.~(\ref{Det}) of the main text.

Finally, we observe that the remaining two factors in the JPDF from Eq.~(\ref{SM:jpdf-susy}) can be expressed as
\begin{equation}
e^{-i\lambda_1(\vec{k}_1^T\vec{s}+\vec{k}_2^T\vec{t})
   + i\lambda_2(\vec{k}_1^T\vec{t}-\vec{k}_2^T\vec{s})}
  = e^{-\frac{N}{2}[z(\vec{k}^T\vec{v}) - z^*(\vec{k}^\dagger\vec{v}^*)]}
  \quad\mbox{and}\quad
I_{BB} = e^{-\frac{N}{2}[\vec{k}^\dagger\vec{k}
       +(\vec{k}^\dagger\vec{v})(\vec{k}^T\vec{v}^*)]}
\end{equation}
in terms of the complex vectors $\vec{v}$ and $\vec{k}$. Note that the scaling with $N$ introduced for $\vec{k}$ contributes an additional $N$-dependent factor to the integration measure $d\vec{k}_1d\vec{k}_2=(N)^{2N}d^2\vec{k}$. Moreover, since the integration over $\vec{q}$ turns out to depend only on its magnitude $q=|\vec{q}|$, the measure becomes $d^2\vec{q}=V_4q^3 dq$, where $V_4=\frac{\pi^2}{2}$ is the volume of the unit ball in the four-dimensional Cartesian space $(\Re{q}_1,\Im{q}_1,\Re{q}_2,\Im{q}_2)$. Collecting all factors together, we arrive at the final integral representation for the JPDF given by Eq.~(\ref{jpdf-exact}) of the main text. In the next section, we develop the method for performing the $\vec{k}$-integration exactly.

\section{Integration over $\vec{k}$ and Wick's theorem}
\label{SM2}
It is convenient to represent the joint distribution in Eq.~(\ref{jpdf-exact}) in the equivalent form
\begin{equation}\label{SM:jpdf-Z}
\mathcal{P}(z,\vec{v}) = \frac{N^{2N+1}}{8\pi}\delta(\vec{v}^\dagger\vec{v}-1)
  \int_0^{\infty}\!dq\,q^3e^{-\frac{N}{2}q^2} \lim_{|w-z|\to0}
  \frac{\partial^2}{\partial{z}\,\partial{z}^*}
  (q^2+|z|^2)^{N-4} \mathcal{Z}(w,z,\vec{v})\,,
\end{equation}
where we have introduced an auxiliary complex variable $w$ to allow swapping the order of the $z$-derivatives and the $\vec{k}$-integration (with the limit $w\to{z}$ to be taken at the end). The generating function $\mathcal{Z}(w,z,\vec{v})$ is defined by
\begin{align}\label{SM:Z}
\mathcal{Z}(w,z,\vec{v}) & \equiv \int\frac{d^2\vec{k}}{(2\pi)^{2N}}
  e^{-\frac{N}{2}\mathcal{L}_{\vec{k}}(w,\vec{v})} D_q(z,\vec{v},\vec{k})
 \\ \nonumber
 & = \int\frac{d^2\vec{k}}{(2\pi)^{2N}}
  e^{-\frac{N}{2}[\vec{k}^\dagger\vec{k}
       + (\vec{k}^\dagger\vec{v})(\vec{k}^T\vec{v}^*)
       + w(\vec{k}^T\vec{v}) - w^*(\vec{k}^\dagger\vec{v}^*)]}
  \det\left[(q^2+|z|^2)\mathbb{I}_4
  - \begin{pmatrix}zA & qB\\ qB^* & -z^*A^*\end{pmatrix}\right]\,.
\end{align}
This representation offers two key advantages. First, performing the $\vec{k}$-integration upfront will drastically reduce the complexity of the expression to be differentiated, as the only remaining invariants involving the vector $\vec{v}$ are $\vec{v}^\dagger\vec{v}=1$ (fixed by normalization) and $\vec{v}^T\vec{v}$. The latter is related to the non-orthogonality overlap. Second, and more importantly, this integration can be carried out exactly because the action depends quadratically on $\vec{k}$, enabling the use of Wick's theorem, which we discuss next.

\subsection{Normal coordinates and diagrammatic rules}

The most efficient way to apply Wick’s theorem in our setting is to first transform the quadratic form $\mathcal{L}_{\vec{k}}(w,\vec{v})$ into normal coordinates and then establish the diagrammatic rules for the second-order correlators in these variables. To achieve this, we first begin by linearizing the mixed term $(\vec{k}^\dagger\vec{v})(\vec{k}^T\vec{v}^*)$ in the exponent of Eq.~(\ref{SM:Z}). This is done making use of the decoupling identity (\ref{SM:decouple}) again and introducing an auxiliary integration over a complex variable $\xi$, \begin{equation}
  e^{-\frac{N}{2}(\vec{k}^\dagger\vec{v})(\vec{k}^T\vec{v}^*)}  =
  \frac{N}{2\pi} \int d^2\xi\,e^{-\frac{N}{2}[\xi\xi^*
  + + i\xi(\vec{k}^\dagger\vec{v}) + i\xi^*(\vec{k}^T\vec{v}^*)]}\,,
\end{equation}
which leads to
\begin{equation}
  e^{-\frac{N}{2}\mathcal{L}_{\vec{k}}(w,\vec{v})}  =
  \frac{N}{2\pi} \int d^2\xi\,e^{-\frac{N}{2} [ \vec{k}^\dagger\vec{k}
  + (w\vec{v}^T + i\xi^*\vec{v}^\dagger)\vec{k}
  - \vec{k}^\dagger(w^*\vec{v}^* - i\xi\vec{v})
  + \xi\xi^*]}\,.
\end{equation}
Centralizing the quadratic form that appears above proceeds in two steps. First, we apply a complex vector shift
\begin{equation}
  \vec{k} \to \vec{r} = \vec{k} - (w^*\vec{v}^*-i\xi \vec{v})
  \qquad\mbox{and}\qquad
  \vec{k}^\dagger\to\vec{r}^\dagger = \vec{k}^\dagger + (w\vec{v}^T+i\xi^*\vec{v}^\dagger)
\end{equation}
to render the quadratic dependence normal in the new coordinates $(\vec{r},\vec{r}^\dagger)$. Next, we centralize the remaining quadratic function in the variables $(\xi, \xi^*)$ by applying  a complex scalar shift
\begin{equation}\textstyle
  \xi \to \zeta = \xi + \frac{i}{2}w^*(\vec{v}^T\vec{v})^*
  \qquad\mbox{and}\qquad
  \xi^*\to\zeta^* = \xi^*-\frac{i}{2}w(\vec{v}^T\vec{v})\,.
\end{equation}
Combined, these steps yield the following transformation to normal coordinates:
\begin{equation}\label{SM:normal}
  \textstyle
  \vec{k} = \vec{r} - i\zeta\vec{v}
          + w^*[\vec{v}^*-\frac{(\vec{v}^T\vec{v})^*}{2}\vec{v}]
  \qquad\mbox{and}\qquad
  \vec{k}^\dagger = \vec{r}^\dagger - i\zeta^*\vec{v}^\dagger
          - w[\vec{v}^T-\frac{(\vec{v}^T\vec{v})}{2}\vec{v}^\dagger]\,.
\end{equation}
As a result, the generating function (\ref{SM:Z}) takes the following attractive representation:
\begin{equation}\label{SM:Z_int}
  \mathcal{Z}(w,z,\vec{v}) =
  e^{-\frac{N}{2}|w|^2(1-\frac{1}{2}|\vec{v}^T\vec{v}|^2)}
  \frac{N}{2\pi}\iint\frac{d^2\vec{r}\,d^2\zeta}{(2\pi)^{2N}}
    e^{-\frac{N}{2}\vec{r}^\dagger\vec{r}} e^{-N|\zeta|^2}
     D_q[z,\vec{v},\vec{k}(\vec{r},\zeta)]\,.
\end{equation}
This representation simplifies the task of performing the $\vec{k}$-integration substantially and reduces it to averaging over two independent -- one vector ($\vec{r}$) and one scalar ($\zeta$) -- uncorrelated Gaussian fields.

The pre-exponential factor $D_q$ is a forth-order polynomial in these variables. Therefore, we can apply Wick's theorem directly, once the second-order correlators are specified. To this end, it is convenient to represent expression (\ref{SM:Z_int}) as
\begin{equation}
  \mathcal{Z}(w,z,\vec{v}) = \frac{1}{2(2\pi N)^N}
  e^{-\frac{N}{2}|w|^2(1-\frac{1}{2}|\vec{v}^T\vec{v}|^2)}
  \aver{\aver{D_q}}\,,
\end{equation}
where we have introduced the notation $\aver{\aver{\cdots}}$ to denote avereging over the Gaussian fields
\begin{equation}\label{SM:Z_aver}
   \aver{\aver{\cdots}} = \frac{N^{N+1}}{\pi(2\pi)^N}
   \iint d^2\vec{r}\,d^2\zeta\,
   e^{-\frac{N}{2}\vec{r}^\dagger\vec{r}} e^{-N|\zeta|^2} (\cdots),
\end{equation}
with the normalization constant chosen to satisfy $\aver{\aver{1}}=1$. This implies zero mean for the components
\begin{equation}
  \aver{\aver{r_i}} = \aver{\aver{\zeta}} = 0
   \qquad\mbox{and}\qquad
  \aver{\aver{r_i\zeta}} = \aver{\aver{r_i}}\aver{\aver{\zeta}} = 0\,.
\end{equation}
The only nonzero contributions to (\ref{SM:Z_aver}) arise from  the correlators of the second and fourth order, which are given by
 \begin{equation}
  \aver{\aver{|\zeta|^2}} = \frac{1}{N}\,, \qquad
  \aver{\aver{r^*_ir_j}} = \frac{2}{N}\delta_{ij}
\end{equation}
and
\begin{equation}
  \aver{\aver{|\zeta|^4}} = \frac{2}{N^2}\,, \qquad
  \aver{\aver{r^*_i r^*_j r_k r_l}} =
  \frac{4}{N^2}(\delta_{ik}\delta_{jl}+\delta_{il}\delta_{jk})\,.
\end{equation}
These relations allow us to formulate a set of substitution rules to efficiently implement the integration in Eq.~(\ref{SM:Z_int}).

\subsection{Substitution rules for $\vec{k}$-integration}

To evaluate $\mathcal{Z}(w,z,\vec{v})$, we first require an explicit form for the determinant $D_q$. This determinant is obtained through a straightforward yet lengthy calculation and can be expressed in the following polynomial form:
\begin{equation}\label{SM:D_q}
  D_q = R^4 + C_3R^3 + C_2R^2 + C_1R + C_0,
  \qquad\mbox{with}\quad R = q^2+|z|^2.
\end{equation}
The four coefficients $C_n$, $n=0,\ldots,3$ are, in turn, given by ($4{-}n$)-th order polynomials in $\vec{k}$ as follows
\begin{subequations}\label{SM:detcoeff}
\begin{align}
C_3 &= -2z(\vec{k}^T\vec{v}) +2z^*(\vec{k}^\dagger\vec{v}^*),  \\
C_2 &= -2q^2(\vec{k}^\dagger\vec{k}+|\vec{k}^\dagger\vec{v}|^2)
    - 4|z|^2|\vec{k}^T\vec{v}|^2
    + z^2\bigl[(\vec{k}^T\vec{v})^2-(\vec{k}^T\vec{k})(\vec{v}^T\vec{v})\bigr]
    + \mbox{c.c.},\\
 C_1 &= 2q^2\Bigl(z\bigl[
        (\vec{k}^\dagger\vec{k})(\vec{k}^T\vec{v})
      - (\vec{k}^T\vec{k})(\vec{k}^\dagger\vec{v})
      + |\vec{k}^\dagger\vec{v}|^2(\vec{k}^T\vec{v})
      - (\vec{k}^\dagger\vec{k})(\vec{k}^\dagger\vec{v}^*)(\vec{v}^T\vec{v})
      \bigr]-\mbox{c.c.} \Bigr)  \nonumber \\
    &\quad +2|z|^2\Bigl(z\bigl[ (\vec{k}^T\vec{v})|\vec{k}^T\vec{v}|^2
      - (\vec{k}^T\vec{k})(\vec{k}^\dagger\vec{v}^*)(\vec{v}^T\vec{v})
      \bigr]-\mbox{c.c.}\Bigr),  \\
 C_0 &= q^4\Bigl(\vec{k}^\dagger\vec{k}-|\vec{k}^\dagger\vec{v}|^2\Bigr)^2
      + q^2|z|^2\Bigl(|\vec{k}^T\vec{k}|^2
        +2(\vec{k}^\dagger\vec{k})|\vec{k}^T\vec{v}|^2
        +2|\vec{k}^\dagger\vec{v}|^2|\vec{k}^T\vec{v}|^2
        -2(\vec{k}^T\vec{k})(\vec{k}^\dagger\vec{v}^*)(\vec{k}^\dagger\vec{v})
        \nonumber \\
   &\quad +\bigl[ (\vec{k}^T\vec{k})(\vec{k}^\dagger\vec{v})^2
       -2(\vec{k}^\dagger\vec{k})(\vec{k}^\dagger\vec{v})(\vec{k}^T\vec{v})
       \bigr](\vec{v}^T\vec{v})^* +\mbox{c.c.}\Bigr)
     + |z|^4\Bigl|(\vec{k}^\dagger\vec{v})^2-(\vec{k}^T\vec{k})(\vec{v}^T\vec{v})\Bigr|^2.
\end{align}
\end{subequations}

The next step is to represent these coefficients in terms of the normal coordinates $(\vec{r},\zeta)$, using the transformation (\ref{SM:normal}). Applying the diagrammatic rules derived above then yields expressions for the average coefficients $\aver{\aver{C_n}}$ and, ultimately, for the average determinant $\aver{\aver{D_q}}$. For instance, one readily finds:
\begin{equation}\textstyle
  \daver{\vec{k}} = w^*[\vec{v}^*-\frac{(\vec{v}^T\vec{v})^*}{2}\vec{v}]
  \qquad\mbox{and}\qquad
  \daver{\vec{k}^\dagger} = - w[\vec{v}^T-\frac{(\vec{v}^T\vec{v})}{2}\vec{v}^\dagger]\,.
\end{equation}
Noting that $\vec{v}^T\vec{v}^*=(\vec{v}^\dagger\vec{v})^*=1$, the first (linear in $\vec{k}$) average coefficient $\aver{\aver{C_3}}$ is obtained exactly as
\begin{equation}\textstyle
  \daver{C_3} = -2(zw^*+z^*w)\nu, \qquad\mbox{with}\quad
  \nu \equiv 1-\frac{1}{2}|\vec{v}^T\vec{v}|^2.
\end{equation}
For $\daver{C_2}$ and the remaining coefficients, which involve higher-order terms in $\vec{k}$, it is convenient to average over $\zeta$ first, reducing the algebraic complexity before averaging over $\vec{r}$. To this end, we note that the vector $\vec{r}$ enters as a fourth-order polynomial involving scalar products $(\vec{r}^T\vec{a})$, where $\vec{a}\in\{\vec{r},\vec{r}^*,\vec{v},\vec{v}^*\}$. It is therefore convenient, for practical purposes, to reformulate the diagrammatic rules in an equivalent form involving these scalar products. All nonzero second-order averages are then given by
\begin{equation}\label{SM:rules2}\textstyle
  \daver{\vec{r}^\dagger\vec{r}} = 2 \qquad\mbox{and}\qquad
  \daver{(\vec{r}^\dagger\vec{a})(\vec{r}^T\vec{b}} = \frac{2}{N}(\vec{a}^T\vec{b}).
\end{equation}
The nonzero averages of the fourth order are given by
\def\norm{\bigl({\textstyle\frac{2}{N}}\bigr)^2}
\begin{equation}\label{SM:rules4diag}
  \daver{(\vec{r}^\dagger\vec{r})^2} = N(N+1)\norm \qquad\mbox{and}\qquad
  \daver{|\vec{r}^T\vec{r}|^2} = 2N\norm
\end{equation}
for the `diagonal' terms, and by
\begin{subequations}
\begin{align}\label{SM:rules4cross}
  &\daver{(\vec{r}^T\vec{r})(\vec{r}^\dagger\vec{a})^2} = 2(\vec{a}^T\vec{a}) \norm \,,\\
  &\daver{(\vec{r}^T\vec{r})(\vec{r}^\dagger\vec{a})(\vec{r}^\dagger\vec{b})}
  = 2(\vec{a}^T\vec{b})\norm \,,\\
  &\daver{(\vec{r}^\dagger\vec{r})(\vec{r}^\dagger\vec{a})(\vec{r}^T\vec{b})}
  = (N+1)(\vec{a}^T\vec{b}) \norm \,,\\
  &\daver{(\vec{r}^\dagger\vec{a})^2(\vec{r}^T\vec{b})^2}
  = 2(\vec{a}^T\vec{b})^2 \norm \,,\\
  &\daver{(\vec{r}^\dagger\vec{a})(\vec{r}^\dagger\vec{b})(\vec{r}^T\vec{a})(\vec{r}^T\vec{b})}
  = [(\vec{a}^T\vec{b})(\vec{a}^T\vec{b})+(\vec{a}^T\vec{a})(\vec{b}^T\vec{b})]\norm \,,
\end{align}
\end{subequations}
for `cross-diagonal' terms, where now $\vec{a},\vec{b} \in \{\vec{v},\vec{v}^*\}$. In this formulation, the $\vec{k}$-integration is reduced to averaging the Gaussian vector field $\vec{r}$, which can be performed efficiently and transparently. For example,  the next (quadratic) coefficient $\daver{C_2}$ is
\begin{equation}\textstyle
  \daver{C_2} = \frac{2}{N}\left[
   \frac{N}{2}\Bigl(2q^2(|w|^2\nu-2) + [(zw^*)^2+(z^*w)^2](2\nu-1) \Bigr)
   - 4|z|^2\nu\bigl(1-\frac{N}{2}|w|^2\nu\bigr) \right].
\end{equation}
The computation of the last two terms $\daver{C_1}$ and $\daver{C_0}$ is more involved but can be carried out efficiently making use of computer algebra systems like \emph{Mathematica} to implement the substitution rules specified above. This procedure yields the final expression for $\daver{D_q}$ and, consequently, for the generating function $\mathcal{Z}(w,z,\vec{v})$ from (\ref{SM:Z_aver}). Substituting this result into (\ref{SM:jpdf-Z}), performing the integration over $q$, and taking the $2D$-Laplacian gives our final result (\ref{jpdf-exact}) presented in the main text.


\begin{thebibliography}{92}%
\makeatletter
\providecommand \@ifxundefined [1]{%
 \@ifx{#1\undefined}
}%
\providecommand \@ifnum [1]{%
 \ifnum #1\expandafter \@firstoftwo
 \else \expandafter \@secondoftwo
 \fi
}%
\providecommand \@ifx [1]{%
 \ifx #1\expandafter \@firstoftwo
 \else \expandafter \@secondoftwo
 \fi
}%
\providecommand \natexlab [1]{#1}%
\providecommand \enquote  [1]{``#1''}%
\providecommand \bibnamefont  [1]{#1}%
\providecommand \bibfnamefont [1]{#1}%
\providecommand \citenamefont [1]{#1}%
\providecommand \href@noop [0]{\@secondoftwo}%
\providecommand \href [0]{\begingroup \@sanitize@url \@href}%
\providecommand \@href[1]{\@@startlink{#1}\@@href}%
\providecommand \@@href[1]{\endgroup#1\@@endlink}%
\providecommand \@sanitize@url [0]{\catcode `\\12\catcode `\$12\catcode
  `\&12\catcode `\#12\catcode `\^12\catcode `\_12\catcode `\%12\relax}%
\providecommand \@@startlink[1]{}%
\providecommand \@@endlink[0]{}%
\providecommand \url  [0]{\begingroup\@sanitize@url \@url }%
\providecommand \@url [1]{\endgroup\@href {#1}{\urlprefix }}%
\providecommand \urlprefix  [0]{URL }%
\providecommand \Eprint [0]{\href }%
\providecommand \doibase [0]{https://doi.org/}%
\providecommand \selectlanguage [0]{\@gobble}%
\providecommand \bibinfo  [0]{\@secondoftwo}%
\providecommand \bibfield  [0]{\@secondoftwo}%
\providecommand \translation [1]{[#1]}%
\providecommand \BibitemOpen [0]{}%
\providecommand \bibitemStop [0]{}%
\providecommand \bibitemNoStop [0]{.\EOS\space}%
\providecommand \EOS [0]{\spacefactor3000\relax}%
\providecommand \BibitemShut  [1]{\csname bibitem#1\endcsname}%
\let\auto@bib@innerbib\@empty
\bibitem [{\citenamefont {Ashida}\ \emph {et~al.}(2020)\citenamefont {Ashida},
  \citenamefont {Gong},\ and\ \citenamefont {Ueda}}]{ashida2020non}%
  \BibitemOpen
  \bibfield  {author} {\bibinfo {author} {\bibfnamefont {Y.}~\bibnamefont
  {Ashida}}, \bibinfo {author} {\bibfnamefont {Z.}~\bibnamefont {Gong}},\ and\
  \bibinfo {author} {\bibfnamefont {M.}~\bibnamefont {Ueda}},\ }\bibfield
  {title} {\bibinfo {title} {Non-{H}ermitian physics},\ }\href
  {https://doi.org/10.1080/00018732.2021.1876991} {\bibfield  {journal}
  {\bibinfo  {journal} {Adv. Phys.}\ }\textbf {\bibinfo {volume} {69}},\
  \bibinfo {pages} {249} (\bibinfo {year} {2020})}\BibitemShut {NoStop}%
\bibitem [{\citenamefont {Grobe}\ \emph {et~al.}(1988)\citenamefont {Grobe},
  \citenamefont {Haake},\ and\ \citenamefont {Sommers}}]{grobe1988quantum}%
  \BibitemOpen
  \bibfield  {author} {\bibinfo {author} {\bibfnamefont {R.}~\bibnamefont
  {Grobe}}, \bibinfo {author} {\bibfnamefont {F.}~\bibnamefont {Haake}},\ and\
  \bibinfo {author} {\bibfnamefont {H.-J.}\ \bibnamefont {Sommers}},\
  }\bibfield  {title} {\bibinfo {title} {Quantum distinction of regular and
  chaotic dissipative motion},\ }\href
  {https://doi.org/10.1103/PhysRevLett.61.1899} {\bibfield  {journal} {\bibinfo
   {journal} {Phys. Rev. Lett.}\ }\textbf {\bibinfo {volume} {61}},\ \bibinfo
  {pages} {1899} (\bibinfo {year} {1988})}\BibitemShut {NoStop}%
\bibitem [{\citenamefont {Akemann}\ \emph {et~al.}(2019)\citenamefont
  {Akemann}, \citenamefont {Kieburg}, \citenamefont {Mielke},\ and\
  \citenamefont {Prosen}}]{akemann2019universal}%
  \BibitemOpen
  \bibfield  {author} {\bibinfo {author} {\bibfnamefont {G.}~\bibnamefont
  {Akemann}}, \bibinfo {author} {\bibfnamefont {M.}~\bibnamefont {Kieburg}},
  \bibinfo {author} {\bibfnamefont {A.}~\bibnamefont {Mielke}},\ and\ \bibinfo
  {author} {\bibfnamefont {T.}~\bibnamefont {Prosen}},\ }\bibfield  {title}
  {\bibinfo {title} {Universal signature from integrability to chaos in
  dissipative open quantum systems},\ }\href
  {https://doi.org/10.1103/PhysRevLett.123.254101} {\bibfield  {journal}
  {\bibinfo  {journal} {Phys. Rev. Lett.}\ }\textbf {\bibinfo {volume} {123}},\
  \bibinfo {pages} {254101} (\bibinfo {year} {2019})}\BibitemShut {NoStop}%
\bibitem [{\citenamefont {Denisov}\ \emph {et~al.}(2019)\citenamefont
  {Denisov}, \citenamefont {Laptyeva}, \citenamefont {Tarnowski}, \citenamefont
  {Chru{\'s}ci{\'n}ski},\ and\ \citenamefont
  {{\.Z}yczkowski}}]{denisov2019universal}%
  \BibitemOpen
  \bibfield  {author} {\bibinfo {author} {\bibfnamefont {S.}~\bibnamefont
  {Denisov}}, \bibinfo {author} {\bibfnamefont {T.}~\bibnamefont {Laptyeva}},
  \bibinfo {author} {\bibfnamefont {W.}~\bibnamefont {Tarnowski}}, \bibinfo
  {author} {\bibfnamefont {D.}~\bibnamefont {Chru{\'s}ci{\'n}ski}},\ and\
  \bibinfo {author} {\bibfnamefont {K.}~\bibnamefont {{\.Z}yczkowski}},\
  }\bibfield  {title} {\bibinfo {title} {Universal spectra of random {L}indblad
  operators},\ }\href {https://doi.org/10.1103/PhysRevLett.123.140403}
  {\bibfield  {journal} {\bibinfo  {journal} {Phys. Rev. Lett.}\ }\textbf
  {\bibinfo {volume} {123}},\ \bibinfo {pages} {140403} (\bibinfo {year}
  {2019})}\BibitemShut {NoStop}%
\bibitem [{\citenamefont {Jaiswal}\ \emph {et~al.}(2019)\citenamefont
  {Jaiswal}, \citenamefont {Pandey},\ and\ \citenamefont
  {Prakash}}]{jaiswal2019universality}%
  \BibitemOpen
  \bibfield  {author} {\bibinfo {author} {\bibfnamefont {A.~B.}\ \bibnamefont
  {Jaiswal}}, \bibinfo {author} {\bibfnamefont {A.}~\bibnamefont {Pandey}},\
  and\ \bibinfo {author} {\bibfnamefont {R.}~\bibnamefont {Prakash}},\
  }\bibfield  {title} {\bibinfo {title} {Universality classes of quantum
  chaotic dissipative systems},\ }\href
  {https://doi.org/10.1209/0295-5075/127/30004} {\bibfield  {journal} {\bibinfo
   {journal} {Europhys. Lett}\ }\textbf {\bibinfo {volume} {127}},\ \bibinfo
  {pages} {30004} (\bibinfo {year} {2019})}\BibitemShut {NoStop}%
\bibitem [{\citenamefont {Can}(2019)}]{can2019random}%
  \BibitemOpen
  \bibfield  {author} {\bibinfo {author} {\bibfnamefont {T.}~\bibnamefont
  {Can}},\ }\bibfield  {title} {\bibinfo {title} {Random {L}indblad dynamics},\
  }\href {https://doi.org/10.1088/1751-8121/ab4d26} {\bibfield  {journal}
  {\bibinfo  {journal} {J. Phys. A: Math. Theor.}\ }\textbf {\bibinfo {volume}
  {52}},\ \bibinfo {pages} {485302} (\bibinfo {year} {2019})}\BibitemShut
  {NoStop}%
\bibitem [{\citenamefont {Hamazaki}\ \emph {et~al.}(2019)\citenamefont
  {Hamazaki}, \citenamefont {Kawabata},\ and\ \citenamefont
  {Ueda}}]{hamazaki2019non}%
  \BibitemOpen
  \bibfield  {author} {\bibinfo {author} {\bibfnamefont {R.}~\bibnamefont
  {Hamazaki}}, \bibinfo {author} {\bibfnamefont {K.}~\bibnamefont {Kawabata}},\
  and\ \bibinfo {author} {\bibfnamefont {M.}~\bibnamefont {Ueda}},\ }\bibfield
  {title} {\bibinfo {title} {Non-{H}ermitian many-body localization},\ }\href
  {https://doi.org/10.1103/PhysRevLett.123.090603} {\bibfield  {journal}
  {\bibinfo  {journal} {Phys. Rev. Lett.}\ }\textbf {\bibinfo {volume} {123}},\
  \bibinfo {pages} {090603} (\bibinfo {year} {2019})}\BibitemShut {NoStop}%
\bibitem [{\citenamefont {S{\'a}}\ \emph
  {et~al.}(2020{\natexlab{a}})\citenamefont {S{\'a}}, \citenamefont {Ribeiro},\
  and\ \citenamefont {Prosen}}]{sa2020spectrala}%
  \BibitemOpen
  \bibfield  {author} {\bibinfo {author} {\bibfnamefont {L.}~\bibnamefont
  {S{\'a}}}, \bibinfo {author} {\bibfnamefont {P.}~\bibnamefont {Ribeiro}},\
  and\ \bibinfo {author} {\bibfnamefont {T.}~\bibnamefont {Prosen}},\
  }\bibfield  {title} {\bibinfo {title} {Spectral and steady-state properties
  of random {L}iouvillians},\ }\href {https://doi.org/10.1088/1751-8121/ab9337}
  {\bibfield  {journal} {\bibinfo  {journal} {J. Phys. A: Math. Theor.}\
  }\textbf {\bibinfo {volume} {53}},\ \bibinfo {pages} {305303} (\bibinfo
  {year} {2020}{\natexlab{a}})}\BibitemShut {NoStop}%
\bibitem [{\citenamefont {S{\'a}}\ \emph
  {et~al.}(2020{\natexlab{b}})\citenamefont {S{\'a}}, \citenamefont {Ribeiro},
  \citenamefont {Can},\ and\ \citenamefont {Prosen}}]{sa2020spectralb}%
  \BibitemOpen
  \bibfield  {author} {\bibinfo {author} {\bibfnamefont {L.}~\bibnamefont
  {S{\'a}}}, \bibinfo {author} {\bibfnamefont {P.}~\bibnamefont {Ribeiro}},
  \bibinfo {author} {\bibfnamefont {T.}~\bibnamefont {Can}},\ and\ \bibinfo
  {author} {\bibfnamefont {T.}~\bibnamefont {Prosen}},\ }\bibfield  {title}
  {\bibinfo {title} {Spectral transitions and universal steady states in random
  {K}raus maps and circuits},\ }\href
  {https://doi.org/10.1103/PhysRevB.102.134310} {\bibfield  {journal} {\bibinfo
   {journal} {Phys. Rev. B}\ }\textbf {\bibinfo {volume} {102}},\ \bibinfo
  {pages} {134310} (\bibinfo {year} {2020}{\natexlab{b}})}\BibitemShut
  {NoStop}%
\bibitem [{\citenamefont {S{\'a}}\ \emph
  {et~al.}(2020{\natexlab{c}})\citenamefont {S{\'a}}, \citenamefont {Ribeiro},\
  and\ \citenamefont {Prosen}}]{sa2020complex}%
  \BibitemOpen
  \bibfield  {author} {\bibinfo {author} {\bibfnamefont {L.}~\bibnamefont
  {S{\'a}}}, \bibinfo {author} {\bibfnamefont {P.}~\bibnamefont {Ribeiro}},\
  and\ \bibinfo {author} {\bibfnamefont {T.}~\bibnamefont {Prosen}},\
  }\bibfield  {title} {\bibinfo {title} {Complex spacing ratios: A signature of
  dissipative quantum chaos},\ }\href
  {https://doi.org/10.1103/PhysRevX.10.021019} {\bibfield  {journal} {\bibinfo
  {journal} {Phys. Rev. X}\ }\textbf {\bibinfo {volume} {10}},\ \bibinfo
  {pages} {021019} (\bibinfo {year} {2020}{\natexlab{c}})}\BibitemShut
  {NoStop}%
\bibitem [{\citenamefont {Wang}\ \emph {et~al.}(2020)\citenamefont {Wang},
  \citenamefont {Piazza},\ and\ \citenamefont {Luitz}}]{wang2020hierarchy}%
  \BibitemOpen
  \bibfield  {author} {\bibinfo {author} {\bibfnamefont {K.}~\bibnamefont
  {Wang}}, \bibinfo {author} {\bibfnamefont {F.}~\bibnamefont {Piazza}},\ and\
  \bibinfo {author} {\bibfnamefont {D.~J.}\ \bibnamefont {Luitz}},\ }\bibfield
  {title} {\bibinfo {title} {Hierarchy of relaxation timescales in local random
  {L}iouvillians},\ }\href {https://doi.org/10.1103/PhysRevLett.124.100604}
  {\bibfield  {journal} {\bibinfo  {journal} {Phys. Rev. Lett.}\ }\textbf
  {\bibinfo {volume} {124}},\ \bibinfo {pages} {100604} (\bibinfo {year}
  {2020})}\BibitemShut {NoStop}%
\bibitem [{\citenamefont {Li}\ \emph {et~al.}(2021)\citenamefont {Li},
  \citenamefont {Prosen},\ and\ \citenamefont {Chan}}]{li2021spectral}%
  \BibitemOpen
  \bibfield  {author} {\bibinfo {author} {\bibfnamefont {J.}~\bibnamefont
  {Li}}, \bibinfo {author} {\bibfnamefont {T.}~\bibnamefont {Prosen}},\ and\
  \bibinfo {author} {\bibfnamefont {A.}~\bibnamefont {Chan}},\ }\bibfield
  {title} {\bibinfo {title} {Spectral statistics of non-{H}ermitian matrices
  and dissipative quantum chaos},\ }\href
  {https://doi.org/10.1103/PhysRevLett.127.170602} {\bibfield  {journal}
  {\bibinfo  {journal} {Phys. Rev. Lett.}\ }\textbf {\bibinfo {volume} {127}},\
  \bibinfo {pages} {170602} (\bibinfo {year} {2021})}\BibitemShut {NoStop}%
\bibitem [{\citenamefont {Tarnowski}\ \emph {et~al.}(2021)\citenamefont
  {Tarnowski}, \citenamefont {Yusipov}, \citenamefont {Laptyeva}, \citenamefont
  {Denisov}, \citenamefont {Chru{\'s}ci{\'n}ski},\ and\ \citenamefont
  {{\.Z}yczkowski}}]{tarnowski2021random}%
  \BibitemOpen
  \bibfield  {author} {\bibinfo {author} {\bibfnamefont {W.}~\bibnamefont
  {Tarnowski}}, \bibinfo {author} {\bibfnamefont {I.}~\bibnamefont {Yusipov}},
  \bibinfo {author} {\bibfnamefont {T.}~\bibnamefont {Laptyeva}}, \bibinfo
  {author} {\bibfnamefont {S.}~\bibnamefont {Denisov}}, \bibinfo {author}
  {\bibfnamefont {D.}~\bibnamefont {Chru{\'s}ci{\'n}ski}},\ and\ \bibinfo
  {author} {\bibfnamefont {K.}~\bibnamefont {{\.Z}yczkowski}},\ }\bibfield
  {title} {\bibinfo {title} {Random generators of {M}arkovian evolution: A
  quantum-classical transition by superdecoherence},\ }\href
  {https://doi.org/10.1103/PhysRevE.104.034118} {\bibfield  {journal} {\bibinfo
   {journal} {Phys. Rev. E}\ }\textbf {\bibinfo {volume} {104}},\ \bibinfo
  {pages} {034118} (\bibinfo {year} {2021})}\BibitemShut {NoStop}%
\bibitem [{\citenamefont {Lange}\ and\ \citenamefont
  {Timm}(2021)}]{lange2021random}%
  \BibitemOpen
  \bibfield  {author} {\bibinfo {author} {\bibfnamefont {S.}~\bibnamefont
  {Lange}}\ and\ \bibinfo {author} {\bibfnamefont {C.}~\bibnamefont {Timm}},\
  }\bibfield  {title} {\bibinfo {title} {Random-matrix theory for the
  {L}indblad master equation},\ }\href {https://doi.org/10.1063/5.0033486}
  {\bibfield  {journal} {\bibinfo  {journal} {Chaos}\ }\textbf {\bibinfo
  {volume} {31}},\ \bibinfo {pages} {023101} (\bibinfo {year}
  {2021})}\BibitemShut {NoStop}%
\bibitem [{\citenamefont {Garc{\'\i}a-Garc{\'\i}a}\ \emph
  {et~al.}(2022)\citenamefont {Garc{\'\i}a-Garc{\'\i}a}, \citenamefont
  {S{\'a}},\ and\ \citenamefont {Verbaarschot}}]{garcia2022symmetry}%
  \BibitemOpen
  \bibfield  {author} {\bibinfo {author} {\bibfnamefont {A.~M.}\ \bibnamefont
  {Garc{\'\i}a-Garc{\'\i}a}}, \bibinfo {author} {\bibfnamefont
  {L.}~\bibnamefont {S{\'a}}},\ and\ \bibinfo {author} {\bibfnamefont {J.~J.}\
  \bibnamefont {Verbaarschot}},\ }\bibfield  {title} {\bibinfo {title}
  {Symmetry classification and universality in non-{H}ermitian many-body
  quantum chaos by the {S}achdev-{Y}e-{K}itaev model},\ }\href
  {https://doi.org/10.1103/PhysRevX.12.021040} {\bibfield  {journal} {\bibinfo
  {journal} {Phys. Rev. X}\ }\textbf {\bibinfo {volume} {12}},\ \bibinfo
  {pages} {021040} (\bibinfo {year} {2022})}\BibitemShut {NoStop}%
\bibitem [{\citenamefont {Kulkarni}\ \emph {et~al.}(2022)\citenamefont
  {Kulkarni}, \citenamefont {Numasawa},\ and\ \citenamefont
  {Ryu}}]{kulkarni2022lindbladian}%
  \BibitemOpen
  \bibfield  {author} {\bibinfo {author} {\bibfnamefont {A.}~\bibnamefont
  {Kulkarni}}, \bibinfo {author} {\bibfnamefont {T.}~\bibnamefont {Numasawa}},\
  and\ \bibinfo {author} {\bibfnamefont {S.}~\bibnamefont {Ryu}},\ }\bibfield
  {title} {\bibinfo {title} {Lindbladian dynamics of the
  {S}achdev-{Y}e-{K}itaev model},\ }\href
  {https://doi.org/10.1103/PhysRevB.106.075138} {\bibfield  {journal} {\bibinfo
   {journal} {Phys. Rev. B}\ }\textbf {\bibinfo {volume} {106}},\ \bibinfo
  {pages} {075138} (\bibinfo {year} {2022})}\BibitemShut {NoStop}%
\bibitem [{\citenamefont {Ghosh}\ \emph {et~al.}(2022)\citenamefont {Ghosh},
  \citenamefont {Gupta},\ and\ \citenamefont {Kulkarni}}]{ghosh2022spectral}%
  \BibitemOpen
  \bibfield  {author} {\bibinfo {author} {\bibfnamefont {S.}~\bibnamefont
  {Ghosh}}, \bibinfo {author} {\bibfnamefont {S.}~\bibnamefont {Gupta}},\ and\
  \bibinfo {author} {\bibfnamefont {M.}~\bibnamefont {Kulkarni}},\ }\bibfield
  {title} {\bibinfo {title} {Spectral properties of disordered interacting
  non-{H}ermitian systems},\ }\href
  {https://doi.org/10.1103/PhysRevB.106.134202} {\bibfield  {journal} {\bibinfo
   {journal} {Phys. Rev. B}\ }\textbf {\bibinfo {volume} {106}},\ \bibinfo
  {pages} {134202} (\bibinfo {year} {2022})}\BibitemShut {NoStop}%
\bibitem [{\citenamefont {Suthar}\ \emph {et~al.}(2022)\citenamefont {Suthar},
  \citenamefont {Wang}, \citenamefont {Huang}, \citenamefont {Jen},\ and\
  \citenamefont {You}}]{suthar2022non}%
  \BibitemOpen
  \bibfield  {author} {\bibinfo {author} {\bibfnamefont {K.}~\bibnamefont
  {Suthar}}, \bibinfo {author} {\bibfnamefont {Y.-C.}\ \bibnamefont {Wang}},
  \bibinfo {author} {\bibfnamefont {Y.-P.}\ \bibnamefont {Huang}}, \bibinfo
  {author} {\bibfnamefont {H.}~\bibnamefont {Jen}},\ and\ \bibinfo {author}
  {\bibfnamefont {J.-S.}\ \bibnamefont {You}},\ }\bibfield  {title} {\bibinfo
  {title} {Non-{H}ermitian many-body localization with open boundaries},\
  }\href {https://doi.org/10.1103/PhysRevB.106.064208} {\bibfield  {journal}
  {\bibinfo  {journal} {Phys. Rev. B}\ }\textbf {\bibinfo {volume} {106}},\
  \bibinfo {pages} {064208} (\bibinfo {year} {2022})}\BibitemShut {NoStop}%
\bibitem [{\citenamefont {Cipolloni}\ and\ \citenamefont
  {Kudler-Flam}(2023{\natexlab{a}})}]{cipolloni2023entanglement}%
  \BibitemOpen
  \bibfield  {author} {\bibinfo {author} {\bibfnamefont {G.}~\bibnamefont
  {Cipolloni}}\ and\ \bibinfo {author} {\bibfnamefont {J.}~\bibnamefont
  {Kudler-Flam}},\ }\bibfield  {title} {\bibinfo {title} {Entanglement entropy
  of non-{H}ermitian eigenstates and the {G}inibre ensemble},\ }\href
  {https://doi.org/10.1103/PhysRevLett.130.010401} {\bibfield  {journal}
  {\bibinfo  {journal} {Phys. Rev. Lett.}\ }\textbf {\bibinfo {volume} {130}},\
  \bibinfo {pages} {010401} (\bibinfo {year} {2023}{\natexlab{a}})}\BibitemShut
  {NoStop}%
\bibitem [{\citenamefont {S{\'a}}\ \emph {et~al.}(2023)\citenamefont {S{\'a}},
  \citenamefont {Ribeiro},\ and\ \citenamefont {Prosen}}]{sa2023symmetry}%
  \BibitemOpen
  \bibfield  {author} {\bibinfo {author} {\bibfnamefont {L.}~\bibnamefont
  {S{\'a}}}, \bibinfo {author} {\bibfnamefont {P.}~\bibnamefont {Ribeiro}},\
  and\ \bibinfo {author} {\bibfnamefont {T.}~\bibnamefont {Prosen}},\
  }\bibfield  {title} {\bibinfo {title} {Symmetry classification of many-body
  {L}indbladians: Tenfold way and beyond},\ }\href
  {https://doi.org/10.1103/PhysRevX.13.031019} {\bibfield  {journal} {\bibinfo
  {journal} {Phys. Rev. X}\ }\textbf {\bibinfo {volume} {13}},\ \bibinfo
  {pages} {031019} (\bibinfo {year} {2023})}\BibitemShut {NoStop}%
\bibitem [{\citenamefont {De~Tomasi}\ and\ \citenamefont
  {Khaymovich}(2022)}]{de2022non}%
  \BibitemOpen
  \bibfield  {author} {\bibinfo {author} {\bibfnamefont {G.}~\bibnamefont
  {De~Tomasi}}\ and\ \bibinfo {author} {\bibfnamefont {I.~M.}\ \bibnamefont
  {Khaymovich}},\ }\bibfield  {title} {\bibinfo {title} {non-{H}ermitian
  {R}osenzweig-{P}orter random-matrix ensemble: Obstruction to the fractal
  phase},\ }\href {https://doi.org/10.1103/PhysRevB.106.094204} {\bibfield
  {journal} {\bibinfo  {journal} {Phys. Rev. B}\ }\textbf {\bibinfo {volume}
  {106}},\ \bibinfo {pages} {094204} (\bibinfo {year} {2022})}\BibitemShut
  {NoStop}%
\bibitem [{\citenamefont {De~Tomasi}\ and\ \citenamefont
  {Khaymovich}(2023)}]{de2023non}%
  \BibitemOpen
  \bibfield  {author} {\bibinfo {author} {\bibfnamefont {G.}~\bibnamefont
  {De~Tomasi}}\ and\ \bibinfo {author} {\bibfnamefont {I.~M.}\ \bibnamefont
  {Khaymovich}},\ }\bibfield  {title} {\bibinfo {title} {Non-{H}ermiticity
  induces localization: Good and bad resonances in power-law random banded
  matrices},\ }\href {https://doi.org/10.1103/PhysRevB.108.L180202} {\bibfield
  {journal} {\bibinfo  {journal} {Phys. Rev. B}\ }\textbf {\bibinfo {volume}
  {108}},\ \bibinfo {pages} {L180202} (\bibinfo {year} {2023})}\BibitemShut
  {NoStop}%
\bibitem [{\citenamefont {Ghosh}\ \emph {et~al.}(2023)\citenamefont {Ghosh},
  \citenamefont {Kulkarni},\ and\ \citenamefont {Roy}}]{ghosh2023eigenvector}%
  \BibitemOpen
  \bibfield  {author} {\bibinfo {author} {\bibfnamefont {S.}~\bibnamefont
  {Ghosh}}, \bibinfo {author} {\bibfnamefont {M.}~\bibnamefont {Kulkarni}},\
  and\ \bibinfo {author} {\bibfnamefont {S.}~\bibnamefont {Roy}},\ }\bibfield
  {title} {\bibinfo {title} {Eigenvector correlations across the localisation
  transition in non-{H}ermitian power-law banded random matrices},\ }\href
  {https://doi.org/10.1103/PhysRevB.108.L060201} {\bibfield  {journal}
  {\bibinfo  {journal} {Phys. Rev. B}\ }\textbf {\bibinfo {volume} {108}},\
  \bibinfo {pages} {L060201} (\bibinfo {year} {2023})}\BibitemShut {NoStop}%
\bibitem [{\citenamefont {Orgad}\ \emph {et~al.}(2024)\citenamefont {Orgad},
  \citenamefont {Oganesyan},\ and\ \citenamefont
  {Gopalakrishnan}}]{orgad2024dynamical}%
  \BibitemOpen
  \bibfield  {author} {\bibinfo {author} {\bibfnamefont {D.}~\bibnamefont
  {Orgad}}, \bibinfo {author} {\bibfnamefont {V.}~\bibnamefont {Oganesyan}},\
  and\ \bibinfo {author} {\bibfnamefont {S.}~\bibnamefont {Gopalakrishnan}},\
  }\bibfield  {title} {\bibinfo {title} {Dynamical transitions from slow to
  fast relaxation in random open quantum systems},\ }\href
  {https://doi.org/10.1103/PhysRevLett.132.040403} {\bibfield  {journal}
  {\bibinfo  {journal} {Phys. Rev. Lett.}\ }\textbf {\bibinfo {volume} {132}},\
  \bibinfo {pages} {040403} (\bibinfo {year} {2024})}\BibitemShut {NoStop}%
\bibitem [{\citenamefont {Jisha}\ and\ \citenamefont
  {Prakash}(2024)}]{jisha2024universality}%
  \BibitemOpen
  \bibfield  {author} {\bibinfo {author} {\bibfnamefont {C.}~\bibnamefont
  {Jisha}}\ and\ \bibinfo {author} {\bibfnamefont {R.}~\bibnamefont
  {Prakash}},\ }\bibfield  {title} {\bibinfo {title} {Universality of spectral
  fluctuations in open quantum chaotic systems},\ }\href
  {https://doi.org/10.1209/0295-5075/ad2c35} {\bibfield  {journal} {\bibinfo
  {journal} {EPL}\ }\textbf {\bibinfo {volume} {146}},\ \bibinfo {pages}
  {11001} (\bibinfo {year} {2024})}\BibitemShut {NoStop}%
\bibitem [{\citenamefont {Richter}\ \emph {et~al.}(2025)\citenamefont
  {Richter}, \citenamefont {S{\'a}},\ and\ \citenamefont
  {Haque}}]{richter2025integrability}%
  \BibitemOpen
  \bibfield  {author} {\bibinfo {author} {\bibfnamefont {J.}~\bibnamefont
  {Richter}}, \bibinfo {author} {\bibfnamefont {L.}~\bibnamefont {S{\'a}}},\
  and\ \bibinfo {author} {\bibfnamefont {M.}~\bibnamefont {Haque}},\ }\bibfield
   {title} {\bibinfo {title} {Integrability versus chaos in the steady state of
  many-body open quantum systems},\ }\href
  {https://doi.org/10.1103/PhysRevE.111.064103} {\bibfield  {journal} {\bibinfo
   {journal} {Phys. Rev. E}\ }\textbf {\bibinfo {volume} {111}},\ \bibinfo
  {pages} {064103} (\bibinfo {year} {2025})}\BibitemShut {NoStop}%
\bibitem [{\citenamefont {Almeida}\ \emph {et~al.}(2025)\citenamefont
  {Almeida}, \citenamefont {Ribeiro}, \citenamefont {Haque},\ and\
  \citenamefont {S{\'a}}}]{almeida2025universality}%
  \BibitemOpen
  \bibfield  {author} {\bibinfo {author} {\bibfnamefont {G.}~\bibnamefont
  {Almeida}}, \bibinfo {author} {\bibfnamefont {P.}~\bibnamefont {Ribeiro}},
  \bibinfo {author} {\bibfnamefont {M.}~\bibnamefont {Haque}},\ and\ \bibinfo
  {author} {\bibfnamefont {L.}~\bibnamefont {S{\'a}}},\ }\bibfield  {title}
  {\bibinfo {title} {Universality, robustness, and limits of the eigenstate
  thermalization hypothesis in open quantum systems},\ }\href@noop {}
  {\bibfield  {journal} {\bibinfo  {journal} {arXiv preprint arXiv:2504.10261}\
  } (\bibinfo {year} {2025})}\BibitemShut {NoStop}%
\bibitem [{\citenamefont {Rufo}\ \emph {et~al.}(2025)\citenamefont {Rufo},
  \citenamefont {Rufo}, \citenamefont {Ribeiro},\ and\ \citenamefont
  {Chesi}}]{rufo2025quantum}%
  \BibitemOpen
  \bibfield  {author} {\bibinfo {author} {\bibfnamefont {G.}~\bibnamefont
  {Rufo}}, \bibinfo {author} {\bibfnamefont {S.}~\bibnamefont {Rufo}}, \bibinfo
  {author} {\bibfnamefont {P.}~\bibnamefont {Ribeiro}},\ and\ \bibinfo {author}
  {\bibfnamefont {S.}~\bibnamefont {Chesi}},\ }\bibfield  {title} {\bibinfo
  {title} {Quantum and semi-classical signatures of dissipative chaos in the
  steady state},\ }\href@noop {} {\bibfield  {journal} {\bibinfo  {journal}
  {arXiv preprint arXiv:2506.14961}\ } (\bibinfo {year} {2025})}\BibitemShut
  {NoStop}%
\bibitem [{\citenamefont {Villase{\~n}or}\ \emph {et~al.}(2025)\citenamefont
  {Villase{\~n}or}, \citenamefont {Yan}, \citenamefont {Orel},\ and\
  \citenamefont {Robnik}}]{villasenor2025correspondence}%
  \BibitemOpen
  \bibfield  {author} {\bibinfo {author} {\bibfnamefont {D.}~\bibnamefont
  {Villase{\~n}or}}, \bibinfo {author} {\bibfnamefont {H.}~\bibnamefont {Yan}},
  \bibinfo {author} {\bibfnamefont {M.}~\bibnamefont {Orel}},\ and\ \bibinfo
  {author} {\bibfnamefont {M.}~\bibnamefont {Robnik}},\ }\bibfield  {title}
  {\bibinfo {title} {Correspondence principle, dissipation, and {G}inibre
  ensemble},\ }\href@noop {} {\bibfield  {journal} {\bibinfo  {journal} {arXiv
  preprint arXiv:2507.18704}\ } (\bibinfo {year} {2025})}\BibitemShut {NoStop}%
\bibitem [{\citenamefont {Wold}\ \emph {et~al.}(2025)\citenamefont {Wold},
  \citenamefont {Zhu}, \citenamefont {Jin}, \citenamefont {Zhu}, \citenamefont
  {Bao}, \citenamefont {Zhong}, \citenamefont {Shen}, \citenamefont {Zhang},
  \citenamefont {Li}, \citenamefont {Wang} \emph
  {et~al.}}]{wold2025experimental}%
  \BibitemOpen
  \bibfield  {author} {\bibinfo {author} {\bibfnamefont {K.}~\bibnamefont
  {Wold}}, \bibinfo {author} {\bibfnamefont {Z.}~\bibnamefont {Zhu}}, \bibinfo
  {author} {\bibfnamefont {F.}~\bibnamefont {Jin}}, \bibinfo {author}
  {\bibfnamefont {X.}~\bibnamefont {Zhu}}, \bibinfo {author} {\bibfnamefont
  {Z.}~\bibnamefont {Bao}}, \bibinfo {author} {\bibfnamefont {J.}~\bibnamefont
  {Zhong}}, \bibinfo {author} {\bibfnamefont {F.}~\bibnamefont {Shen}},
  \bibinfo {author} {\bibfnamefont {P.}~\bibnamefont {Zhang}}, \bibinfo
  {author} {\bibfnamefont {H.}~\bibnamefont {Li}}, \bibinfo {author}
  {\bibfnamefont {Z.}~\bibnamefont {Wang}}, \emph {et~al.},\ }\bibfield
  {title} {\bibinfo {title} {Experimental detection of dissipative quantum
  chaos},\ }\href@noop {} {\bibfield  {journal} {\bibinfo  {journal} {arXiv
  preprint arXiv:2506.04325}\ } (\bibinfo {year} {2025})}\BibitemShut {NoStop}%
\bibitem [{\citenamefont {Chirame}\ and\ \citenamefont
  {Burnell}(2025)}]{chirame2025open}%
  \BibitemOpen
  \bibfield  {author} {\bibinfo {author} {\bibfnamefont {S.}~\bibnamefont
  {Chirame}}\ and\ \bibinfo {author} {\bibfnamefont {F.~J.}\ \bibnamefont
  {Burnell}},\ }\bibfield  {title} {\bibinfo {title} {Open system dynamics in
  local lindbladians with chaotic spectra},\ }\href@noop {} {\bibfield
  {journal} {\bibinfo  {journal} {arXiv preprint arXiv:2510.15193}\ } (\bibinfo
  {year} {2025})}\BibitemShut {NoStop}%
\bibitem [{\citenamefont {Hamazaki}\ \emph {et~al.}(2020)\citenamefont
  {Hamazaki}, \citenamefont {Kawabata}, \citenamefont {Kura},\ and\
  \citenamefont {Ueda}}]{hamazaki2020universality}%
  \BibitemOpen
  \bibfield  {author} {\bibinfo {author} {\bibfnamefont {R.}~\bibnamefont
  {Hamazaki}}, \bibinfo {author} {\bibfnamefont {K.}~\bibnamefont {Kawabata}},
  \bibinfo {author} {\bibfnamefont {N.}~\bibnamefont {Kura}},\ and\ \bibinfo
  {author} {\bibfnamefont {M.}~\bibnamefont {Ueda}},\ }\bibfield  {title}
  {\bibinfo {title} {Universality classes of non-{H}ermitian random matrices},\
  }\href {https://doi.org/10.1103/PhysRevResearch.2.023286} {\bibfield
  {journal} {\bibinfo  {journal} {Phys. Rev. Research}\ }\textbf {\bibinfo
  {volume} {2}},\ \bibinfo {pages} {023286} (\bibinfo {year}
  {2020})}\BibitemShut {NoStop}%
\bibitem [{\citenamefont {Kawabata}\ \emph {et~al.}(2019)\citenamefont
  {Kawabata}, \citenamefont {Shiozaki}, \citenamefont {Ueda},\ and\
  \citenamefont {Sato}}]{Kawabata2019}%
  \BibitemOpen
  \bibfield  {author} {\bibinfo {author} {\bibfnamefont {K.}~\bibnamefont
  {Kawabata}}, \bibinfo {author} {\bibfnamefont {K.}~\bibnamefont {Shiozaki}},
  \bibinfo {author} {\bibfnamefont {M.}~\bibnamefont {Ueda}},\ and\ \bibinfo
  {author} {\bibfnamefont {M.}~\bibnamefont {Sato}},\ }\bibfield  {title}
  {\bibinfo {title} {Symmetry and topology in non-{H}ermitian physics},\ }\href
  {https://doi.org/10.1103/PhysRevX.9.041015} {\bibfield  {journal} {\bibinfo
  {journal} {Phys. Rev. X}\ }\textbf {\bibinfo {volume} {9}},\ \bibinfo {pages}
  {041015} (\bibinfo {year} {2019})}\BibitemShut {NoStop}%
\bibitem [{\citenamefont {Bernard}\ and\ \citenamefont
  {LeClair}(2002)}]{Bernard2002}%
  \BibitemOpen
  \bibfield  {author} {\bibinfo {author} {\bibfnamefont {D.}~\bibnamefont
  {Bernard}}\ and\ \bibinfo {author} {\bibfnamefont {A.}~\bibnamefont
  {LeClair}},\ }\bibinfo {title} {A classification of non-{H}ermitian random
  matrices},\ in\ \href {https://doi.org/10.1007/978-94-010-0514-2_19} {\emph
  {\bibinfo {booktitle} {Statistical Field Theories}}},\ \bibinfo {editor}
  {edited by\ \bibinfo {editor} {\bibfnamefont {A.}~\bibnamefont {Cappelli}}\
  and\ \bibinfo {editor} {\bibfnamefont {G.}~\bibnamefont {Mussardo}}}\
  (\bibinfo  {publisher} {Springer Netherlands},\ \bibinfo {address}
  {Dordrecht},\ \bibinfo {year} {2002})\ pp.\ \bibinfo {pages}
  {207--214}\BibitemShut {NoStop}%
\bibitem [{\citenamefont {Magnea}(2008)}]{Magnea2008}%
  \BibitemOpen
  \bibfield  {author} {\bibinfo {author} {\bibfnamefont {U.}~\bibnamefont
  {Magnea}},\ }\bibfield  {title} {\bibinfo {title} {Random matrices beyond the
  {C}artan classification},\ }\href
  {https://doi.org/10.1088/1751-8113/41/4/045203} {\bibfield  {journal}
  {\bibinfo  {journal} {J. Phys. A: Math. Theor.}\ }\textbf {\bibinfo {volume}
  {41}},\ \bibinfo {pages} {045203} (\bibinfo {year} {2008})}\BibitemShut
  {NoStop}%
\bibitem [{\citenamefont {Akemann}\ \emph
  {et~al.}(2025{\natexlab{a}})\citenamefont {Akemann}, \citenamefont
  {Ayg{\"u}n}, \citenamefont {Kieburg},\ and\ \citenamefont
  {P{\"a}{\ss}ler}}]{akemann2025complex}%
  \BibitemOpen
  \bibfield  {author} {\bibinfo {author} {\bibfnamefont {G.}~\bibnamefont
  {Akemann}}, \bibinfo {author} {\bibfnamefont {N.}~\bibnamefont {Ayg{\"u}n}},
  \bibinfo {author} {\bibfnamefont {M.}~\bibnamefont {Kieburg}},\ and\ \bibinfo
  {author} {\bibfnamefont {P.}~\bibnamefont {P{\"a}{\ss}ler}},\ }\bibfield
  {title} {\bibinfo {title} {Complex symmetric, self-dual, and {G}inibre random
  matrices: analytical results for three classes of bulk and edge statistics},\
  }\href {https://doi.org/10.1088/1751-8121/adbd9d} {\bibfield  {journal}
  {\bibinfo  {journal} {J. Phys. A: Math. Theor.}\ }\textbf {\bibinfo {volume}
  {58}},\ \bibinfo {pages} {125204} (\bibinfo {year}
  {2025}{\natexlab{a}})}\BibitemShut {NoStop}%
\bibitem [{\citenamefont {Byun}\ and\ \citenamefont
  {Forrester}(2025)}]{byun2025progress}%
  \BibitemOpen
  \bibfield  {author} {\bibinfo {author} {\bibfnamefont {S.-S.}\ \bibnamefont
  {Byun}}\ and\ \bibinfo {author} {\bibfnamefont {P.~J.}\ \bibnamefont
  {Forrester}},\ }\href@noop {} {\emph {\bibinfo {title} {Progress on the Study
  of the {G}inibre Ensembles}}}\ (\bibinfo  {publisher} {Springer Nature},\
  \bibinfo {year} {2025})\BibitemShut {NoStop}%
\bibitem [{\citenamefont {Akemann}\ \emph {et~al.}(2022)\citenamefont
  {Akemann}, \citenamefont {Mielke},\ and\ \citenamefont
  {P{\"a}{\ss}ler}}]{akemann2022spacing}%
  \BibitemOpen
  \bibfield  {author} {\bibinfo {author} {\bibfnamefont {G.}~\bibnamefont
  {Akemann}}, \bibinfo {author} {\bibfnamefont {A.}~\bibnamefont {Mielke}},\
  and\ \bibinfo {author} {\bibfnamefont {P.}~\bibnamefont {P{\"a}{\ss}ler}},\
  }\bibfield  {title} {\bibinfo {title} {Spacing distribution in the
  two-dimensional coulomb gas: Surmise and symmetry classes of non-{H}ermitian
  random matrices at noninteger $\beta$},\ }\href
  {https://doi.org/10.1103/PhysRevE.106.014146} {\bibfield  {journal} {\bibinfo
   {journal} {Phys. Rev. E}\ }\textbf {\bibinfo {volume} {106}},\ \bibinfo
  {pages} {014146} (\bibinfo {year} {2022})}\BibitemShut {NoStop}%
\bibitem [{\citenamefont {Akemann}\ \emph
  {et~al.}(2025{\natexlab{b}})\citenamefont {Akemann}, \citenamefont
  {Balducci}, \citenamefont {Chenu}, \citenamefont {P{\"a}{\ss}ler},
  \citenamefont {Roccati},\ and\ \citenamefont {Shir}}]{akemann2025two}%
  \BibitemOpen
  \bibfield  {author} {\bibinfo {author} {\bibfnamefont {G.}~\bibnamefont
  {Akemann}}, \bibinfo {author} {\bibfnamefont {F.}~\bibnamefont {Balducci}},
  \bibinfo {author} {\bibfnamefont {A.}~\bibnamefont {Chenu}}, \bibinfo
  {author} {\bibfnamefont {P.}~\bibnamefont {P{\"a}{\ss}ler}}, \bibinfo
  {author} {\bibfnamefont {F.}~\bibnamefont {Roccati}},\ and\ \bibinfo {author}
  {\bibfnamefont {R.}~\bibnamefont {Shir}},\ }\bibfield  {title} {\bibinfo
  {title} {Two transitions in complex eigenvalue statistics: Hermiticity and
  integrability breaking},\ }\href
  {https://doi.org/10.1103/PhysRevResearch.7.013098} {\bibfield  {journal}
  {\bibinfo  {journal} {Phys. Rev. Research}\ }\textbf {\bibinfo {volume}
  {7}},\ \bibinfo {pages} {013098} (\bibinfo {year}
  {2025}{\natexlab{b}})}\BibitemShut {NoStop}%
\bibitem [{\citenamefont {Kanazawa}\ and\ \citenamefont
  {Wettig}(2021)}]{kanazawa2021new}%
  \BibitemOpen
  \bibfield  {author} {\bibinfo {author} {\bibfnamefont {T.}~\bibnamefont
  {Kanazawa}}\ and\ \bibinfo {author} {\bibfnamefont {T.}~\bibnamefont
  {Wettig}},\ }\bibfield  {title} {\bibinfo {title} {New universality classes
  of the non-{H}ermitian dirac operator in qcd-like theories},\ }\href
  {https://doi.org/10.1103/PhysRevD.104.014509} {\bibfield  {journal} {\bibinfo
   {journal} {Phys. Rev. D}\ }\textbf {\bibinfo {volume} {104}},\ \bibinfo
  {pages} {014509} (\bibinfo {year} {2021})}\BibitemShut {NoStop}%
\bibitem [{\citenamefont {Huang}\ and\ \citenamefont
  {Shklovskii}(2020)}]{huang2020spectral}%
  \BibitemOpen
  \bibfield  {author} {\bibinfo {author} {\bibfnamefont {Y.}~\bibnamefont
  {Huang}}\ and\ \bibinfo {author} {\bibfnamefont {B.}~\bibnamefont
  {Shklovskii}},\ }\bibfield  {title} {\bibinfo {title} {Spectral rigidity of
  non-{H}ermitian symmetric random matrices near the anderson transition},\
  }\href {https://doi.org/10.1103/PhysRevB.102.064212} {\bibfield  {journal}
  {\bibinfo  {journal} {Phys. Rev. B}\ }\textbf {\bibinfo {volume} {102}},\
  \bibinfo {pages} {064212} (\bibinfo {year} {2020})}\BibitemShut {NoStop}%
\bibitem [{\citenamefont {Sommers}\ \emph {et~al.}(1999)\citenamefont
  {Sommers}, \citenamefont {Fyodorov},\ and\ \citenamefont
  {Titov}}]{sommers1999s}%
  \BibitemOpen
  \bibfield  {author} {\bibinfo {author} {\bibfnamefont {H.-J.}\ \bibnamefont
  {Sommers}}, \bibinfo {author} {\bibfnamefont {Y.~V.}\ \bibnamefont
  {Fyodorov}},\ and\ \bibinfo {author} {\bibfnamefont {M.}~\bibnamefont
  {Titov}},\ }\bibfield  {title} {\bibinfo {title} {S-matrix poles for chaotic
  quantum systems as eigenvalues of complex symmetric random matrices: from
  isolated to overlapping resonances},\ }\href
  {https://doi.org/10.1088/0305-4470/32/5/003} {\bibfield  {journal} {\bibinfo
  {journal} {J. Phys. A}\ }\textbf {\bibinfo {volume} {32}},\ \bibinfo {pages}
  {L77} (\bibinfo {year} {1999})}\BibitemShut {NoStop}%
\bibitem [{\citenamefont {Prado}\ \emph {et~al.}(2025)\citenamefont {Prado},
  \citenamefont {Bachelard}, \citenamefont {Kaiser},\ and\ \citenamefont
  {Pinheiro}}]{prado2025nonergodic}%
  \BibitemOpen
  \bibfield  {author} {\bibinfo {author} {\bibfnamefont {M.}~\bibnamefont
  {Prado}}, \bibinfo {author} {\bibfnamefont {R.}~\bibnamefont {Bachelard}},
  \bibinfo {author} {\bibfnamefont {R.}~\bibnamefont {Kaiser}},\ and\ \bibinfo
  {author} {\bibfnamefont {F.~A.}\ \bibnamefont {Pinheiro}},\ }\bibfield
  {title} {\bibinfo {title} {Nonergodic extended phase for waves in three
  dimensions},\ }\href@noop {} {\bibfield  {journal} {\bibinfo  {journal}
  {arXiv:2510.20346}\ } (\bibinfo {year} {2025})}\BibitemShut {NoStop}%
\bibitem [{\citenamefont {Forrester}(2025)}]{forrester2025dualities}%
  \BibitemOpen
  \bibfield  {author} {\bibinfo {author} {\bibfnamefont {P.~J.}\ \bibnamefont
  {Forrester}},\ }\bibfield  {title} {\bibinfo {title} {Dualities for
  characteristic polynomial averages of complex symmetric and self dual
  non-{H}ermitian random matrices},\ }\href
  {https://doi.org/10.1088/1751-8121/adacb8} {\bibfield  {journal} {\bibinfo
  {journal} {J. Phys. A: Math. Theor.}\ }\textbf {\bibinfo {volume} {58}},\
  \bibinfo {pages} {075201} (\bibinfo {year} {2025})}\BibitemShut {NoStop}%
\bibitem [{\citenamefont {Kulkarni}\ \emph {et~al.}(2025)\citenamefont
  {Kulkarni}, \citenamefont {Kawabata},\ and\ \citenamefont
  {Ryu}}]{kulkarni25}%
  \BibitemOpen
  \bibfield  {author} {\bibinfo {author} {\bibfnamefont {A.}~\bibnamefont
  {Kulkarni}}, \bibinfo {author} {\bibfnamefont {K.}~\bibnamefont {Kawabata}},\
  and\ \bibinfo {author} {\bibfnamefont {S.}~\bibnamefont {Ryu}},\ }\bibfield
  {title} {\bibinfo {title} {Non-linear sigma models for non-{H}ermitian random
  matrices in symmetry classes {AI$^\dagger$} and {AII$^\dagger$}},\ }\href
  {https://doi.org/10.1088/1751-8121/adc95f} {\bibfield  {journal} {\bibinfo
  {journal} {J. Phys. A: Math. Theor.}\ }\textbf {\bibinfo {volume} {58}},\
  \bibinfo {pages} {225202} (\bibinfo {year} {2025})}\BibitemShut {NoStop}%
\bibitem [{\citenamefont {Trefethen}\ and\ \citenamefont
  {Embree}(2020)}]{trefethen2020spectra}%
  \BibitemOpen
  \bibfield  {author} {\bibinfo {author} {\bibfnamefont {L.~N.}\ \bibnamefont
  {Trefethen}}\ and\ \bibinfo {author} {\bibfnamefont {M.}~\bibnamefont
  {Embree}},\ }\href@noop {} {\emph {\bibinfo {title} {Spectra and
  pseudospectra: the behavior of nonnormal matrices and operators}}}\ (\bibinfo
   {publisher} {Princeton university press},\ \bibinfo {year}
  {2020})\BibitemShut {NoStop}%
\bibitem [{\citenamefont {Schomerus}\ \emph {et~al.}(2000)\citenamefont
  {Schomerus}, \citenamefont {Frahm}, \citenamefont {Patra},\ and\
  \citenamefont {Beenakker}}]{schomerus2000quantum}%
  \BibitemOpen
  \bibfield  {author} {\bibinfo {author} {\bibfnamefont {H.}~\bibnamefont
  {Schomerus}}, \bibinfo {author} {\bibfnamefont {K.~M.}\ \bibnamefont
  {Frahm}}, \bibinfo {author} {\bibfnamefont {M.}~\bibnamefont {Patra}},\ and\
  \bibinfo {author} {\bibfnamefont {C.~W.~J.}\ \bibnamefont {Beenakker}},\
  }\bibfield  {title} {\bibinfo {title} {Quantum limit of the laser line width
  in chaotic cavities and statistics of residues of scattering matrix poles},\
  }\href {https://doi.org/10.1016/S0378-4371(99)00602-0} {\bibfield  {journal}
  {\bibinfo  {journal} {Physica A}\ }\textbf {\bibinfo {volume} {278}},\
  \bibinfo {pages} {469} (\bibinfo {year} {2000})}\BibitemShut {NoStop}%
\bibitem [{\citenamefont {Savin}\ and\ \citenamefont {Sokolov}(1997)}]{savi97}%
  \BibitemOpen
  \bibfield  {author} {\bibinfo {author} {\bibfnamefont {D.~V.}\ \bibnamefont
  {Savin}}\ and\ \bibinfo {author} {\bibfnamefont {V.~V.}\ \bibnamefont
  {Sokolov}},\ }\bibfield  {title} {\bibinfo {title} {Quantum versus classical
  decay laws in open chaotic systems},\ }\href
  {https://doi.org/10.1103/PhysRevE.56.R4911} {\bibfield  {journal} {\bibinfo
  {journal} {Phys. Rev. E}\ }\textbf {\bibinfo {volume} {56}},\ \bibinfo
  {pages} {R4911} (\bibinfo {year} {1997})}\BibitemShut {NoStop}%
\bibitem [{\citenamefont {Chalker}\ and\ \citenamefont
  {Mehlig}(1998)}]{ChalkerMehlig1998}%
  \BibitemOpen
  \bibfield  {author} {\bibinfo {author} {\bibfnamefont {J.~T.}\ \bibnamefont
  {Chalker}}\ and\ \bibinfo {author} {\bibfnamefont {B.}~\bibnamefont
  {Mehlig}},\ }\bibfield  {title} {\bibinfo {title} {Eigenvector statistics in
  non-{H}ermitian random matrix ensembles},\ }\href
  {https://doi.org/10.1103/PhysRevLett.81.3367} {\bibfield  {journal} {\bibinfo
   {journal} {Phys. Rev. Lett.}\ }\textbf {\bibinfo {volume} {81}},\ \bibinfo
  {pages} {3367} (\bibinfo {year} {1998})}\BibitemShut {NoStop}%
\bibitem [{\citenamefont {Mehlig}\ and\ \citenamefont
  {Chalker}(2000)}]{ChalkerMehlig2000}%
  \BibitemOpen
  \bibfield  {author} {\bibinfo {author} {\bibfnamefont {B.}~\bibnamefont
  {Mehlig}}\ and\ \bibinfo {author} {\bibfnamefont {J.~T.}\ \bibnamefont
  {Chalker}},\ }\bibfield  {title} {\bibinfo {title} {Statistical properties of
  eigenvectors in non-{H}ermitian {G}aussian random matrix ensembles},\ }\href
  {https://doi.org/10.1063/1.533302} {\bibfield  {journal} {\bibinfo  {journal}
  {J. Math. Phys.}\ }\textbf {\bibinfo {volume} {41}},\ \bibinfo {pages} {3233}
  (\bibinfo {year} {2000})}\BibitemShut {NoStop}%
\bibitem [{\citenamefont {Burda}\ \emph
  {et~al.}(2014{\natexlab{a}})\citenamefont {Burda}, \citenamefont {Grela},
  \citenamefont {Nowak}, \citenamefont {Tarnowski},\ and\ \citenamefont
  {Warcho\l}}]{burd14}%
  \BibitemOpen
  \bibfield  {author} {\bibinfo {author} {\bibfnamefont {Z.}~\bibnamefont
  {Burda}}, \bibinfo {author} {\bibfnamefont {J.}~\bibnamefont {Grela}},
  \bibinfo {author} {\bibfnamefont {M.~A.}\ \bibnamefont {Nowak}}, \bibinfo
  {author} {\bibfnamefont {W.}~\bibnamefont {Tarnowski}},\ and\ \bibinfo
  {author} {\bibfnamefont {P.}~\bibnamefont {Warcho\l}},\ }\bibfield  {title}
  {\bibinfo {title} {Dysonian dynamics of the {G}inibre ensemble},\ }\href
  {https://doi.org/10.1103/PhysRevLett.113.104102} {\bibfield  {journal}
  {\bibinfo  {journal} {Phys. Rev. Lett.}\ }\textbf {\bibinfo {volume} {113}},\
  \bibinfo {pages} {104102} (\bibinfo {year} {2014}{\natexlab{a}})}\BibitemShut
  {NoStop}%
\bibitem [{\citenamefont {Erdos}\ \emph {et~al.}(2018)\citenamefont {Erdos},
  \citenamefont {Kru{\"u}ger},\ and\ \citenamefont {Renfrew}}]{erdos2018power}%
  \BibitemOpen
  \bibfield  {author} {\bibinfo {author} {\bibfnamefont {L.}~\bibnamefont
  {Erdos}}, \bibinfo {author} {\bibfnamefont {T.}~\bibnamefont {Kru{\"u}ger}},\
  and\ \bibinfo {author} {\bibfnamefont {D.}~\bibnamefont {Renfrew}},\
  }\bibfield  {title} {\bibinfo {title} {Power law decay for systems of
  randomly coupled differential equations},\ }\href
  {https://doi.org/10.1137/17M1143125} {\bibfield  {journal} {\bibinfo
  {journal} {SIAM J. Math. Anal.}\ }\textbf {\bibinfo {volume} {50}},\ \bibinfo
  {pages} {3271} (\bibinfo {year} {2018})}\BibitemShut {NoStop}%
\bibitem [{\citenamefont {Tarnowski}\ \emph {et~al.}(2020)\citenamefont
  {Tarnowski}, \citenamefont {Neri},\ and\ \citenamefont
  {Vivo}}]{Tarnowski_2020}%
  \BibitemOpen
  \bibfield  {author} {\bibinfo {author} {\bibfnamefont {W.}~\bibnamefont
  {Tarnowski}}, \bibinfo {author} {\bibfnamefont {I.}~\bibnamefont {Neri}},\
  and\ \bibinfo {author} {\bibfnamefont {P.}~\bibnamefont {Vivo}},\ }\bibfield
  {title} {\bibinfo {title} {Universal transient behaviour in large dynamical
  systems on networks},\ }\href
  {https://doi.org/10.1103/PhysRevResearch.2.023333} {\bibfield  {journal}
  {\bibinfo  {journal} {Phys. Rev. Research}\ }\textbf {\bibinfo {volume}
  {2}},\ \bibinfo {pages} {023333} (\bibinfo {year} {2020})}\BibitemShut
  {NoStop}%
\bibitem [{\citenamefont {Gudowska-Nowak}\ \emph {et~al.}(2020)\citenamefont
  {Gudowska-Nowak}, \citenamefont {Nowak}, \citenamefont {Chialvo},
  \citenamefont {Ochab},\ and\ \citenamefont {Tarnowski}}]{Gudowska_neuro}%
  \BibitemOpen
  \bibfield  {author} {\bibinfo {author} {\bibfnamefont {E.}~\bibnamefont
  {Gudowska-Nowak}}, \bibinfo {author} {\bibfnamefont {M.~A.}\ \bibnamefont
  {Nowak}}, \bibinfo {author} {\bibfnamefont {D.~R.}\ \bibnamefont {Chialvo}},
  \bibinfo {author} {\bibfnamefont {J.~K.}\ \bibnamefont {Ochab}},\ and\
  \bibinfo {author} {\bibfnamefont {W.}~\bibnamefont {Tarnowski}},\ }\bibfield
  {title} {\bibinfo {title} {From synaptic interactions to collective dynamics
  in random neuronal networks models: Critical role of eigenvectors and
  transient behavior},\ }\href {https://doi.org/10.1162/neco_a_01253}
  {\bibfield  {journal} {\bibinfo  {journal} {Neural Comput.}\ }\textbf
  {\bibinfo {volume} {32}},\ \bibinfo {pages} {395} (\bibinfo {year}
  {2020})}\BibitemShut {NoStop}%
\bibitem [{\citenamefont {Fyodorov}\ \emph {et~al.}(2025)\citenamefont
  {Fyodorov}, \citenamefont {Gudowska-Nowak}, \citenamefont {Nowak},\ and\
  \citenamefont {Tarnowski}}]{Fyodorov_entropyproduction_2025}%
  \BibitemOpen
  \bibfield  {author} {\bibinfo {author} {\bibfnamefont {Y.~V.}\ \bibnamefont
  {Fyodorov}}, \bibinfo {author} {\bibfnamefont {E.}~\bibnamefont
  {Gudowska-Nowak}}, \bibinfo {author} {\bibfnamefont {M.}~\bibnamefont
  {Nowak}},\ and\ \bibinfo {author} {\bibfnamefont {W.}~\bibnamefont
  {Tarnowski}},\ }\bibfield  {title} {\bibinfo {title} {Nonorthogonal
  eigenvectors, fluctuation-dissipation relations, and entropy production},\
  }\href {https://doi.org/10.1103/PhysRevLett.134.087102} {\bibfield  {journal}
  {\bibinfo  {journal} {Phys. Rev. Lett.}\ }\textbf {\bibinfo {volume} {134}},\
  \bibinfo {pages} {087102} (\bibinfo {year} {2025})}\BibitemShut {NoStop}%
\bibitem [{\citenamefont {Fyodorov}\ and\ \citenamefont
  {Savin}(2012)}]{fyodorovSavin2012}%
  \BibitemOpen
  \bibfield  {author} {\bibinfo {author} {\bibfnamefont {Y.~V.}\ \bibnamefont
  {Fyodorov}}\ and\ \bibinfo {author} {\bibfnamefont {D.~V.}\ \bibnamefont
  {Savin}},\ }\bibfield  {title} {\bibinfo {title} {Statistics of resonance
  width shifts as a signature of eigenfunction nonorthogonality},\ }\href
  {https://doi.org/10.1103/PhysRevLett.108.184101} {\bibfield  {journal}
  {\bibinfo  {journal} {Phys. Rev. Lett.}\ }\textbf {\bibinfo {volume} {108}},\
  \bibinfo {pages} {184101} (\bibinfo {year} {2012})}\BibitemShut {NoStop}%
\bibitem [{\citenamefont {Gros}\ \emph {et~al.}(2014)\citenamefont {Gros},
  \citenamefont {Kuhl}, \citenamefont {Legrand}, \citenamefont {Mortessagne},
  \citenamefont {Richalot},\ and\ \citenamefont {Savin}}]{gros14}%
  \BibitemOpen
  \bibfield  {author} {\bibinfo {author} {\bibfnamefont {J.-B.}\ \bibnamefont
  {Gros}}, \bibinfo {author} {\bibfnamefont {U.}~\bibnamefont {Kuhl}}, \bibinfo
  {author} {\bibfnamefont {O.}~\bibnamefont {Legrand}}, \bibinfo {author}
  {\bibfnamefont {F.}~\bibnamefont {Mortessagne}}, \bibinfo {author}
  {\bibfnamefont {E.}~\bibnamefont {Richalot}},\ and\ \bibinfo {author}
  {\bibfnamefont {D.~V.}\ \bibnamefont {Savin}},\ }\bibfield  {title} {\bibinfo
  {title} {Experimental width shift distribution: A test of nonorthogonality
  for local and global perturbations},\ }\href
  {https://doi.org/10.1103/PhysRevLett.113.224101} {\bibfield  {journal}
  {\bibinfo  {journal} {Phys. Rev. Lett.}\ }\textbf {\bibinfo {volume} {113}},\
  \bibinfo {pages} {224101} (\bibinfo {year} {2014})}\BibitemShut {NoStop}%
\bibitem [{\citenamefont {Davy}\ and\ \citenamefont
  {Genack}(2019)}]{davy2019probing}%
  \BibitemOpen
  \bibfield  {author} {\bibinfo {author} {\bibfnamefont {M.}~\bibnamefont
  {Davy}}\ and\ \bibinfo {author} {\bibfnamefont {A.~Z.}\ \bibnamefont
  {Genack}},\ }\bibfield  {title} {\bibinfo {title} {Probing nonorthogonality
  of eigenfunctions and its impact on transport through open systems},\ }\href
  {https://doi.org/10.1103/PhysRevResearch.1.033026} {\bibfield  {journal}
  {\bibinfo  {journal} {Phys. Rev. Research}\ }\textbf {\bibinfo {volume}
  {1}},\ \bibinfo {pages} {033026} (\bibinfo {year} {2019})}\BibitemShut
  {NoStop}%
\bibitem [{\citenamefont {Fyodorov}\ and\ \citenamefont
  {Osman}(2022)}]{fyodorov2022eigenfunction}%
  \BibitemOpen
  \bibfield  {author} {\bibinfo {author} {\bibfnamefont {Y.~V.}\ \bibnamefont
  {Fyodorov}}\ and\ \bibinfo {author} {\bibfnamefont {M.}~\bibnamefont
  {Osman}},\ }\bibfield  {title} {\bibinfo {title} {Eigenfunction
  non-orthogonality factors and the shape of {CPA}-like dips in a
  single-channel reflection from lossy chaotic cavities},\ }\href
  {https://doi.org/10.1088/1751-8121/ac6717} {\bibfield  {journal} {\bibinfo
  {journal} {J. Phys. A: Math. Theor.}\ }\textbf {\bibinfo {volume} {55}},\
  \bibinfo {pages} {224013} (\bibinfo {year} {2022})}\BibitemShut {NoStop}%
\bibitem [{\citenamefont {Cipolloni}\ and\ \citenamefont
  {Kudler-Flam}(2024)}]{cipolloni2023non}%
  \BibitemOpen
  \bibfield  {author} {\bibinfo {author} {\bibfnamefont {G.}~\bibnamefont
  {Cipolloni}}\ and\ \bibinfo {author} {\bibfnamefont {J.}~\bibnamefont
  {Kudler-Flam}},\ }\bibfield  {title} {\bibinfo {title} {Non-{H}ermitian
  hamiltonians violate the eigenstate thermalization hypothesis},\ }\href
  {https://doi.org/10.1103/PhysRevB.109.L020201} {\bibfield  {journal}
  {\bibinfo  {journal} {Phys. Rev. B}\ }\textbf {\bibinfo {volume} {109}},\
  \bibinfo {pages} {L020201} (\bibinfo {year} {2024})}\BibitemShut {NoStop}%
\bibitem [{\citenamefont {Cipolloni}\ and\ \citenamefont
  {Kudler-Flam}(2023{\natexlab{b}})}]{cipolloni2023entan}%
  \BibitemOpen
  \bibfield  {author} {\bibinfo {author} {\bibfnamefont {G.}~\bibnamefont
  {Cipolloni}}\ and\ \bibinfo {author} {\bibfnamefont {J.}~\bibnamefont
  {Kudler-Flam}},\ }\bibfield  {title} {\bibinfo {title} {Entanglement entropy
  of non-{H}ermitian eigenstates and the {G}inibre ensemble},\ }\href
  {https://doi.org/10.1103/PhysRevLett.130.010401} {\bibfield  {journal}
  {\bibinfo  {journal} {Phys. Rev. Lett.}\ }\textbf {\bibinfo {volume} {130}},\
  \bibinfo {pages} {010401} (\bibinfo {year} {2023}{\natexlab{b}})}\BibitemShut
  {NoStop}%
\bibitem [{\citenamefont {Bao}(2025)}]{bao2025initial}%
  \BibitemOpen
  \bibfield  {author} {\bibinfo {author} {\bibfnamefont {R.}~\bibnamefont
  {Bao}},\ }\bibfield  {title} {\bibinfo {title} {Initial-state typicality in
  quantum relaxation},\ }\href@noop {} {\bibfield  {journal} {\bibinfo
  {journal} {arXiv preprint arXiv:2511.01709}\ } (\bibinfo {year}
  {2025})}\BibitemShut {NoStop}%
\bibitem [{\citenamefont {Janik}\ \emph {et~al.}(1999)\citenamefont {Janik},
  \citenamefont {N{\"o}renberg}, \citenamefont {Nowak}, \citenamefont {Papp},\
  and\ \citenamefont {Zahed}}]{janik1999correlations}%
  \BibitemOpen
  \bibfield  {author} {\bibinfo {author} {\bibfnamefont {R.~A.}\ \bibnamefont
  {Janik}}, \bibinfo {author} {\bibfnamefont {W.}~\bibnamefont
  {N{\"o}renberg}}, \bibinfo {author} {\bibfnamefont {M.~A.}\ \bibnamefont
  {Nowak}}, \bibinfo {author} {\bibfnamefont {G.}~\bibnamefont {Papp}},\ and\
  \bibinfo {author} {\bibfnamefont {I.}~\bibnamefont {Zahed}},\ }\bibfield
  {title} {\bibinfo {title} {Correlations of eigenvectors for non-{H}ermitian
  random-matrix models},\ }\href {https://doi.org/10.1103/PhysRevE.60.2699}
  {\bibfield  {journal} {\bibinfo  {journal} {Phys. Rev. E}\ }\textbf {\bibinfo
  {volume} {60}},\ \bibinfo {pages} {2699} (\bibinfo {year}
  {1999})}\BibitemShut {NoStop}%
\bibitem [{\citenamefont {Fyodorov}\ and\ \citenamefont
  {Mehlig}(2002)}]{fyodorovmehlig2002}%
  \BibitemOpen
  \bibfield  {author} {\bibinfo {author} {\bibfnamefont {Y.~V.}\ \bibnamefont
  {Fyodorov}}\ and\ \bibinfo {author} {\bibfnamefont {B.}~\bibnamefont
  {Mehlig}},\ }\bibfield  {title} {\bibinfo {title} {Statistics of resonances
  and nonorthogonal eigenfunctions in a model for single-channel chaotic
  scattering},\ }\href {https://doi.org/10.1103/PhysRevE.66.045202} {\bibfield
  {journal} {\bibinfo  {journal} {Phys. Rev. E}\ }\textbf {\bibinfo {volume}
  {66}},\ \bibinfo {pages} {045202} (\bibinfo {year} {2002})}\BibitemShut
  {NoStop}%
\bibitem [{\citenamefont {Poli}\ \emph {et~al.}(2009)\citenamefont {Poli},
  \citenamefont {Savin}, \citenamefont {Legrand},\ and\ \citenamefont
  {Mortessagne}}]{poli09b}%
  \BibitemOpen
  \bibfield  {author} {\bibinfo {author} {\bibfnamefont {C.}~\bibnamefont
  {Poli}}, \bibinfo {author} {\bibfnamefont {D.~V.}\ \bibnamefont {Savin}},
  \bibinfo {author} {\bibfnamefont {O.}~\bibnamefont {Legrand}},\ and\ \bibinfo
  {author} {\bibfnamefont {F.}~\bibnamefont {Mortessagne}},\ }\bibfield
  {title} {\bibinfo {title} {Statistics of resonance states in open chaotic
  systems: A perturbative approach},\ }\href
  {https://doi.org/10.1103/PhysRevE.80.046203} {\bibfield  {journal} {\bibinfo
  {journal} {Phys. Rev. E}\ }\textbf {\bibinfo {volume} {80}},\ \bibinfo
  {pages} {046203} (\bibinfo {year} {2009})}\BibitemShut {NoStop}%
\bibitem [{\citenamefont {Burda}\ \emph
  {et~al.}(2014{\natexlab{b}})\citenamefont {Burda}, \citenamefont {Grela},
  \citenamefont {Nowak}, \citenamefont {Tarnowski},\ and\ \citenamefont
  {Warcho{\l}}}]{burda2014dysonian}%
  \BibitemOpen
  \bibfield  {author} {\bibinfo {author} {\bibfnamefont {Z.}~\bibnamefont
  {Burda}}, \bibinfo {author} {\bibfnamefont {J.}~\bibnamefont {Grela}},
  \bibinfo {author} {\bibfnamefont {M.~A.}\ \bibnamefont {Nowak}}, \bibinfo
  {author} {\bibfnamefont {W.}~\bibnamefont {Tarnowski}},\ and\ \bibinfo
  {author} {\bibfnamefont {P.}~\bibnamefont {Warcho{\l}}},\ }\bibfield  {title}
  {\bibinfo {title} {Dysonian dynamics of the {G}inibre ensemble},\ }\href
  {https://doi.org/10.1103/PhysRevLett.113.104102} {\bibfield  {journal}
  {\bibinfo  {journal} {Phys. Rev. Lett.}\ }\textbf {\bibinfo {volume} {113}},\
  \bibinfo {pages} {104102} (\bibinfo {year} {2014}{\natexlab{b}})}\BibitemShut
  {NoStop}%
\bibitem [{\citenamefont {Walters}\ and\ \citenamefont
  {Starr}(2015)}]{walters2015note}%
  \BibitemOpen
  \bibfield  {author} {\bibinfo {author} {\bibfnamefont {M.}~\bibnamefont
  {Walters}}\ and\ \bibinfo {author} {\bibfnamefont {S.}~\bibnamefont
  {Starr}},\ }\bibfield  {title} {\bibinfo {title} {A note on mixed matrix
  moments for the complex {G}inibre ensemble},\ }\bibfield  {journal} {\bibinfo
   {journal} {J. Math. Phys.}\ }\textbf {\bibinfo {volume} {56}},\ \href
  {https://doi.org/10.1063/1.4904451} {10.1063/1.4904451} (\bibinfo {year}
  {2015})\BibitemShut {NoStop}%
\bibitem [{\citenamefont {Belinschi}\ \emph {et~al.}(2017)\citenamefont
  {Belinschi}, \citenamefont {Nowak}, \citenamefont {Speicher},\ and\
  \citenamefont {Tarnowski}}]{belinschi2017squared}%
  \BibitemOpen
  \bibfield  {author} {\bibinfo {author} {\bibfnamefont {S.}~\bibnamefont
  {Belinschi}}, \bibinfo {author} {\bibfnamefont {M.~A.}\ \bibnamefont
  {Nowak}}, \bibinfo {author} {\bibfnamefont {R.}~\bibnamefont {Speicher}},\
  and\ \bibinfo {author} {\bibfnamefont {W.}~\bibnamefont {Tarnowski}},\
  }\bibfield  {title} {\bibinfo {title} {Squared eigenvalue condition numbers
  and eigenvector correlations from the single ring theorem},\ }\href
  {https://doi.org/10.1088/1751-8121/aa5451} {\bibfield  {journal} {\bibinfo
  {journal} {J. Phys. A: Math. Theor.}\ }\textbf {\bibinfo {volume} {50}},\
  \bibinfo {pages} {105204} (\bibinfo {year} {2017})}\BibitemShut {NoStop}%
\bibitem [{\citenamefont {Bourgade}\ and\ \citenamefont
  {Dubach}(2020)}]{bourgadedoubach2020}%
  \BibitemOpen
  \bibfield  {author} {\bibinfo {author} {\bibfnamefont {P.}~\bibnamefont
  {Bourgade}}\ and\ \bibinfo {author} {\bibfnamefont {G.}~\bibnamefont
  {Dubach}},\ }\bibfield  {title} {\bibinfo {title} {The distribution of
  overlaps between eigenvectors of {G}inibre matrices},\ }\href
  {https://doi.org/10.1007/s00440-019-00953-x} {\bibfield  {journal} {\bibinfo
  {journal} {Probabil. Theor. Related Fields}\ }\textbf {\bibinfo {volume}
  {177}},\ \bibinfo {pages} {397} (\bibinfo {year} {2020})}\BibitemShut
  {NoStop}%
\bibitem [{\citenamefont {Fyodorov}(2018)}]{fyodorov2018CMP}%
  \BibitemOpen
  \bibfield  {author} {\bibinfo {author} {\bibfnamefont {Y.~V.}\ \bibnamefont
  {Fyodorov}},\ }\bibfield  {title} {\bibinfo {title} {On statistics of
  bi-orthogonal eigenvectors in real and complex {G}inibre ensembles: combining
  partial schur decomposition with supersymmetry},\ }\href
  {https://doi.org/10.1007/s00220-018-3163-3} {\bibfield  {journal} {\bibinfo
  {journal} {Comm. Math. Phys.}\ }\textbf {\bibinfo {volume} {363}},\ \bibinfo
  {pages} {579} (\bibinfo {year} {2018})}\BibitemShut {NoStop}%
\bibitem [{\citenamefont {Akemann}\ \emph
  {et~al.}(2020{\natexlab{a}})\citenamefont {Akemann}, \citenamefont
  {Förster},\ and\ \citenamefont {Kieburg}}]{Akemann2020quatern}%
  \BibitemOpen
  \bibfield  {author} {\bibinfo {author} {\bibfnamefont {G.}~\bibnamefont
  {Akemann}}, \bibinfo {author} {\bibfnamefont {Y.-P.}\ \bibnamefont
  {Förster}},\ and\ \bibinfo {author} {\bibfnamefont {M.}~\bibnamefont
  {Kieburg}},\ }\bibfield  {title} {\bibinfo {title} {Universal eigenvector
  correlations in quaternionic {G}inibre ensembles},\ }\href
  {https://doi.org/10.1088/1751-8121/ab766e} {\bibfield  {journal} {\bibinfo
  {journal} {J. Phys. A: Math. Theor.}\ }\textbf {\bibinfo {volume} {53}},\
  \bibinfo {pages} {145201} (\bibinfo {year} {2020}{\natexlab{a}})}\BibitemShut
  {NoStop}%
\bibitem [{\citenamefont {Dubach}(2021{\natexlab{a}})}]{dubach2021a}%
  \BibitemOpen
  \bibfield  {author} {\bibinfo {author} {\bibfnamefont {G.}~\bibnamefont
  {Dubach}},\ }\bibfield  {title} {\bibinfo {title} {Symmetries of the
  quaternionic {G}inibre ensemble},\ }\href
  {https://doi.org/10.1142/S2010326321500131} {\bibfield  {journal} {\bibinfo
  {journal} {Random Matrices: Theory Appl.}\ }\textbf {\bibinfo {volume}
  {10}},\ \bibinfo {pages} {2150013} (\bibinfo {year}
  {2021}{\natexlab{a}})}\BibitemShut {NoStop}%
\bibitem [{\citenamefont {W{\"u}rfel}\ \emph
  {et~al.}(2024{\natexlab{a}})\citenamefont {W{\"u}rfel}, \citenamefont
  {Crumpton},\ and\ \citenamefont {Fyodorov}}]{WFC1}%
  \BibitemOpen
  \bibfield  {author} {\bibinfo {author} {\bibfnamefont {T.~R.}\ \bibnamefont
  {W{\"u}rfel}}, \bibinfo {author} {\bibfnamefont {M.~J.}\ \bibnamefont
  {Crumpton}},\ and\ \bibinfo {author} {\bibfnamefont {Y.~V.}\ \bibnamefont
  {Fyodorov}},\ }\bibfield  {title} {\bibinfo {title} {Mean left-right
  eigenvector self-overlap in the real {G}inibre ensemble},\ }\href
  {https://doi.org/10.1142/S2010326324500175} {\bibfield  {journal} {\bibinfo
  {journal} {Random Matrices: Theory Appl.}\ }\textbf {\bibinfo {volume}
  {13}},\ \bibinfo {pages} {2450017} (\bibinfo {year}
  {2024}{\natexlab{a}})}\BibitemShut {NoStop}%
\bibitem [{\citenamefont {Akemann}\ \emph
  {et~al.}(2020{\natexlab{b}})\citenamefont {Akemann}, \citenamefont {Tribe},
  \citenamefont {Tsareas},\ and\ \citenamefont
  {Zaboronski}}]{akemann2020determinantal}%
  \BibitemOpen
  \bibfield  {author} {\bibinfo {author} {\bibfnamefont {G.}~\bibnamefont
  {Akemann}}, \bibinfo {author} {\bibfnamefont {R.}~\bibnamefont {Tribe}},
  \bibinfo {author} {\bibfnamefont {A.}~\bibnamefont {Tsareas}},\ and\ \bibinfo
  {author} {\bibfnamefont {O.}~\bibnamefont {Zaboronski}},\ }\bibfield  {title}
  {\bibinfo {title} {On the determinantal structure of conditional overlaps for
  the complex {G}inibre ensemble},\ }\href
  {https://doi.org/10.1142/S201032632050015X} {\bibfield  {journal} {\bibinfo
  {journal} {Random Matrices: Theory Appl.}\ }\textbf {\bibinfo {volume} {9}},\
  \bibinfo {pages} {2050015} (\bibinfo {year}
  {2020}{\natexlab{b}})}\BibitemShut {NoStop}%
\bibitem [{\citenamefont {Akemann}\ \emph
  {et~al.}(2020{\natexlab{c}})\citenamefont {Akemann}, \citenamefont {Tribe},
  \citenamefont {Tsareas},\ and\ \citenamefont {Zaboronski}}]{akemann2020appa}%
  \BibitemOpen
  \bibfield  {author} {\bibinfo {author} {\bibfnamefont {G.}~\bibnamefont
  {Akemann}}, \bibinfo {author} {\bibfnamefont {R.}~\bibnamefont {Tribe}},
  \bibinfo {author} {\bibfnamefont {A.}~\bibnamefont {Tsareas}},\ and\ \bibinfo
  {author} {\bibfnamefont {O.}~\bibnamefont {Zaboronski}},\ }\bibfield  {title}
  {\bibinfo {title} {Determinantal structure and bulk universality of
  conditional overlaps in the complex {G}inibre ensemble},\ }\href
  {https://doi.org/10.5506/APhysPolB.51.1611} {\bibfield  {journal} {\bibinfo
  {journal} {Acta Phys. Pol.}\ }\textbf {\bibinfo {volume} {51}},\ \bibinfo
  {pages} {1611} (\bibinfo {year} {2020}{\natexlab{c}})}\BibitemShut {NoStop}%
\bibitem [{\citenamefont {Akemann}\ \emph
  {et~al.}(2020{\natexlab{d}})\citenamefont {Akemann}, \citenamefont
  {F{\"o}rster},\ and\ \citenamefont {Kieburg}}]{akemann2020universal}%
  \BibitemOpen
  \bibfield  {author} {\bibinfo {author} {\bibfnamefont {G.}~\bibnamefont
  {Akemann}}, \bibinfo {author} {\bibfnamefont {Y.-P.}\ \bibnamefont
  {F{\"o}rster}},\ and\ \bibinfo {author} {\bibfnamefont {M.}~\bibnamefont
  {Kieburg}},\ }\bibfield  {title} {\bibinfo {title} {Universal eigenvector
  correlations in quaternionic {G}inibre ensembles},\ }\href
  {https://doi.org/10.1088/1751-8121/ab766e} {\bibfield  {journal} {\bibinfo
  {journal} {J. Phys. A: Math. Theor.}\ }\textbf {\bibinfo {volume} {53}},\
  \bibinfo {pages} {145201} (\bibinfo {year} {2020}{\natexlab{d}})}\BibitemShut
  {NoStop}%
\bibitem [{\citenamefont {Dubach}(2021{\natexlab{b}})}]{dubach2021b}%
  \BibitemOpen
  \bibfield  {author} {\bibinfo {author} {\bibfnamefont {G.}~\bibnamefont
  {Dubach}},\ }\bibfield  {title} {\bibinfo {title} {On eigenvector statistics
  in the spherical and truncated unitary ensembles},\ }\href
  {https://doi.org/10.1214/21-EJP686} {\bibfield  {journal} {\bibinfo
  {journal} {Electron. J. Prob.}\ }\textbf {\bibinfo {volume} {26}},\ \bibinfo
  {pages} {1} (\bibinfo {year} {2021}{\natexlab{b}})}\BibitemShut {NoStop}%
\bibitem [{\citenamefont {Fyodorov}\ and\ \citenamefont
  {Tarnowski}(2021)}]{fyodorov2021condition}%
  \BibitemOpen
  \bibfield  {author} {\bibinfo {author} {\bibfnamefont {Y.~V.}\ \bibnamefont
  {Fyodorov}}\ and\ \bibinfo {author} {\bibfnamefont {W.}~\bibnamefont
  {Tarnowski}},\ }\bibfield  {title} {\bibinfo {title} {Condition numbers for
  real eigenvalues in the real elliptic {G}aussian ensemble},\ }\href
  {https://doi.org/10.1007/s00023-020-00967-5} {\bibfield  {journal} {\bibinfo
  {journal} {Annales Henri Poincar{\'e}}\ }\textbf {\bibinfo {volume} {22}},\
  \bibinfo {pages} {309} (\bibinfo {year} {2021})}\BibitemShut {NoStop}%
\bibitem [{\citenamefont {Osman}(2024)}]{osman2024universality}%
  \BibitemOpen
  \bibfield  {author} {\bibinfo {author} {\bibfnamefont {M.}~\bibnamefont
  {Osman}},\ }\bibfield  {title} {\bibinfo {title} {Universality for diagonal
  eigenvector overlaps of non-{H}ermitian random matrices},\ }\href@noop {}
  {\bibfield  {journal} {\bibinfo  {journal} {arXiv preprint arXiv:2409.16144}\
  } (\bibinfo {year} {2024})}\BibitemShut {NoStop}%
\bibitem [{\citenamefont {Cipolloni}\ \emph {et~al.}(2024)\citenamefont
  {Cipolloni}, \citenamefont {Erd{\H{o}}s}, \citenamefont {Henheik},\ and\
  \citenamefont {Schr{\"o}der}}]{cipolloni2024optimala}%
  \BibitemOpen
  \bibfield  {author} {\bibinfo {author} {\bibfnamefont {G.}~\bibnamefont
  {Cipolloni}}, \bibinfo {author} {\bibfnamefont {L.}~\bibnamefont
  {Erd{\H{o}}s}}, \bibinfo {author} {\bibfnamefont {J.}~\bibnamefont
  {Henheik}},\ and\ \bibinfo {author} {\bibfnamefont {D.}~\bibnamefont
  {Schr{\"o}der}},\ }\bibfield  {title} {\bibinfo {title} {Optimal lower bound
  on eigenvector overlaps for non-{H}ermitian random matrices},\ }\href
  {https://doi.org/10.1016/j.jfa.2024.110495} {\bibfield  {journal} {\bibinfo
  {journal} {J. Funct. Anal.}\ }\textbf {\bibinfo {volume} {287}},\ \bibinfo
  {pages} {110495} (\bibinfo {year} {2024})}\BibitemShut {NoStop}%
\bibitem [{\citenamefont {Cipolloni}\ \emph {et~al.}(2026)\citenamefont
  {Cipolloni}, \citenamefont {Erd{\H{o}}s},\ and\ \citenamefont
  {Xu}}]{cipolloni2025optimal}%
  \BibitemOpen
  \bibfield  {author} {\bibinfo {author} {\bibfnamefont {G.}~\bibnamefont
  {Cipolloni}}, \bibinfo {author} {\bibfnamefont {L.}~\bibnamefont
  {Erd{\H{o}}s}},\ and\ \bibinfo {author} {\bibfnamefont {Y.}~\bibnamefont
  {Xu}},\ }\bibfield  {title} {\bibinfo {title} {Optimal decay of eigenvector
  overlap for non-{H}ermitian random matrices},\ }\href
  {https://doi.org/10.1016/j.jfa.2025.111180} {\bibfield  {journal} {\bibinfo
  {journal} {J. Funct. Anal.}\ }\textbf {\bibinfo {volume} {290}},\ \bibinfo
  {pages} {111180} (\bibinfo {year} {2026})}\BibitemShut {NoStop}%
\bibitem [{\citenamefont {Tarnowski}(2024)}]{tarnowski2024condition}%
  \BibitemOpen
  \bibfield  {author} {\bibinfo {author} {\bibfnamefont {W.}~\bibnamefont
  {Tarnowski}},\ }\bibfield  {title} {\bibinfo {title} {Condition numbers for
  real eigenvalues of real elliptic ensemble: weak non-normality at the edge},\
  }\href {https://doi.org/10.1088/1751-8121/ad523b} {\bibfield  {journal}
  {\bibinfo  {journal} {J. Phys. A: Math. Theor.}\ }\textbf {\bibinfo {volume}
  {57}},\ \bibinfo {pages} {255204} (\bibinfo {year} {2024})}\BibitemShut
  {NoStop}%
\bibitem [{\citenamefont {Zhang}(2024)}]{zhang2024mean}%
  \BibitemOpen
  \bibfield  {author} {\bibinfo {author} {\bibfnamefont {L.}~\bibnamefont
  {Zhang}},\ }\bibfield  {title} {\bibinfo {title} {Mean eigenvector
  self-overlap in deformed complex {G}inibre ensemble},\ }\href@noop {}
  {\bibfield  {journal} {\bibinfo  {journal} {arXiv preprint arXiv:2407.09163}\
  } (\bibinfo {year} {2024})}\BibitemShut {NoStop}%
\bibitem [{\citenamefont {W{\"u}rfel}\ \emph
  {et~al.}(2024{\natexlab{b}})\citenamefont {W{\"u}rfel}, \citenamefont
  {Crumpton},\ and\ \citenamefont {Fyodorov}}]{wurfel2024mean}%
  \BibitemOpen
  \bibfield  {author} {\bibinfo {author} {\bibfnamefont {T.~R.}\ \bibnamefont
  {W{\"u}rfel}}, \bibinfo {author} {\bibfnamefont {M.~J.}\ \bibnamefont
  {Crumpton}},\ and\ \bibinfo {author} {\bibfnamefont {Y.~V.}\ \bibnamefont
  {Fyodorov}},\ }\bibfield  {title} {\bibinfo {title} {Mean left-right
  eigenvector self-overlap in the real {G}inibre ensemble},\ }\href
  {https://doi.org/10.1142/S2010326324500175} {\bibfield  {journal} {\bibinfo
  {journal} {Random Matrices: Theory Appl.}\ }\textbf {\bibinfo {volume}
  {13}},\ \bibinfo {pages} {2450017} (\bibinfo {year}
  {2024}{\natexlab{b}})}\BibitemShut {NoStop}%
\bibitem [{\citenamefont {Crumpton}\ \emph {et~al.}(2025)\citenamefont
  {Crumpton}, \citenamefont {Fyodorov},\ and\ \citenamefont
  {W{\"u}rfel}}]{crumpton2025mean}%
  \BibitemOpen
  \bibfield  {author} {\bibinfo {author} {\bibfnamefont {M.~J.}\ \bibnamefont
  {Crumpton}}, \bibinfo {author} {\bibfnamefont {Y.~V.}\ \bibnamefont
  {Fyodorov}},\ and\ \bibinfo {author} {\bibfnamefont {T.~R.}\ \bibnamefont
  {W{\"u}rfel}},\ }\bibfield  {title} {\bibinfo {title} {Mean eigenvector
  self-overlap in the real and complex elliptic {G}inibre ensembles at strong
  and weak non-hermiticity},\ }\href
  {https://doi.org/10.1007/s00023-024-01530-2} {\bibfield  {journal} {\bibinfo
  {journal} {Annales Henri Poincar{\'e}}\ }\textbf {\bibinfo {volume} {26}},\
  \bibinfo {pages} {2069} (\bibinfo {year} {2025})}\BibitemShut {NoStop}%
\bibitem [{\citenamefont {Akemann}\ \emph
  {et~al.}(2025{\natexlab{c}})\citenamefont {Akemann}, \citenamefont {Byun},\
  and\ \citenamefont {Noda}}]{akemann2025pfaffian}%
  \BibitemOpen
  \bibfield  {author} {\bibinfo {author} {\bibfnamefont {G.}~\bibnamefont
  {Akemann}}, \bibinfo {author} {\bibfnamefont {S.-S.}\ \bibnamefont {Byun}},\
  and\ \bibinfo {author} {\bibfnamefont {K.}~\bibnamefont {Noda}},\ }\bibfield
  {title} {\bibinfo {title} {Pfaffian structure of the eigenvector overlap for
  the symplectic {G}inibre ensemble},\ }in\ \href
  {https://doi.org/10.1007/s00023-025-01575-x} {\emph {\bibinfo {booktitle}
  {Annales Henri Poincar{\'e}}}}\ (\bibinfo {organization} {Springer},\
  \bibinfo {year} {2025})\ pp.\ \bibinfo {pages} {1--52}\BibitemShut {NoStop}%
\bibitem [{\citenamefont {Noda}(2025)}]{noda2025determinantal}%
  \BibitemOpen
  \bibfield  {author} {\bibinfo {author} {\bibfnamefont {K.}~\bibnamefont
  {Noda}},\ }\bibfield  {title} {\bibinfo {title} {Determinantal structure of
  the overlaps for induced spherical unitary ensemble},\ }\href
  {https://doi.org/10.1142/S2010326325500121} {\bibfield  {journal} {\bibinfo
  {journal} {Random Matrices: Theory Appl.}\ }\textbf {\bibinfo {volume}
  {14}},\ \bibinfo {pages} {2550012} (\bibinfo {year} {2025})}\BibitemShut
  {NoStop}%
\bibitem [{\citenamefont {Fyodorov}(2025)}]{fyodorov2025kac}%
  \BibitemOpen
  \bibfield  {author} {\bibinfo {author} {\bibfnamefont {Y.~V.}\ \bibnamefont
  {Fyodorov}},\ }\bibfield  {title} {\bibinfo {title} {{Kac-Rice} inspired
  approach to non-{H}ermitian random matrices},\ }\href@noop {} {\bibfield
  {journal} {\bibinfo  {journal} {arXiv preprint arXiv:2506.21058}\ } (\bibinfo
  {year} {2025})}\BibitemShut {NoStop}%
\bibitem [{\citenamefont {Sommers}\ and\ \citenamefont
  {Iida}(1994)}]{sommers1994eigenvector}%
  \BibitemOpen
  \bibfield  {author} {\bibinfo {author} {\bibfnamefont {H.-J.}\ \bibnamefont
  {Sommers}}\ and\ \bibinfo {author} {\bibfnamefont {S.}~\bibnamefont {Iida}},\
  }\bibfield  {title} {\bibinfo {title} {Eigenvector statistics in the
  crossover region between {G}aussian orthogonal and unitary ensembles},\
  }\href {https://doi.org/10.1103/PhysRevE.49.R2513} {\bibfield  {journal}
  {\bibinfo  {journal} {Phys. Rev. E}\ }\textbf {\bibinfo {volume} {49}},\
  \bibinfo {pages} {R2513} (\bibinfo {year} {1994})}\BibitemShut {NoStop}%
\bibitem [{SM()}]{SM}%
  \BibitemOpen
  \href@noop {} {}\bibinfo {note} {See Supplemental Material at [URL will be
  inserted by publisher] for further details of the method and
  calculations.}\BibitemShut {Stop}%
\bibitem [{rad()}]{radius}%
  \BibitemOpen
  \href@noop {} {}\bibinfo {note} {Generally, the case of an arbirary spectral
  radius $a$ is obtained by considering a rescaled random matrix
  $J=\frac{a}{\sqrt{2}}(X+iY)$, which corresponds to replacing $x=N(|z|/a)^2$
  in our main result, Eqs.~(\ref{jpdf-exact}) and (\ref{jpdf-exact_g}), and in
  what follows accordingly.}\BibitemShut {Stop}%
\bibitem [{NIS()}]{NIST}%
  \BibitemOpen
  \href@noop {} {\bibinfo {title} {{NIST Digital Library of Mathematical
  Functions}}},\ \bibinfo {howpublished} {Release 1.2.4 of 2025-03-15,
  available at {\small\url{https://dlmf.nist.gov/}}},\ \bibinfo {note}
  {{F.~W.~J.} Olver, A.~B. Olde Daalhuis, D.~W. Lozier, B.~I. Schneider, R.~F.
  Boisvert, C.~W. Clark, B.~R. Miller, B.~V. Saunders, H.~S. Cohl, and M.~A.
  McClain, eds.}\BibitemShut {Stop}%
\end{thebibliography}
\end{document}